    \magnification=\magstep1
     \hsize=5 in
     \vsize=6.75 in
     \hoffset=.25 in
     \voffset=.5 in
     
     \def\tilde{\widetilde}
     \def\bar{\overline}
     \def\hat{\widehat}
     \def\dsl{\raise .15ex\hbox{/}\kern-.57em\partial}
     \def\Dsl{\,\raise .15ex\hbox{/}\mkern-13.5mu D}
\centerline{{\bf Zeros of the Jimbo, Miwa, Ueno tau function}}
\vskip.2in \par\noindent
\centerline{{\sl John Palmer}}
\centerline{\sl Department of Mathematics}
\centerline{\sl University of Arizona}
\centerline{\sl Tucson, AZ 85721}
\centerline{ \sl email: palmer@math.arizona.edu}
\vskip.2in \par\noindent
\centerline {\bf Abstract}
We introduce a family of local deformations for 
meromorphic connections on ${\bf P}^1$ in the neighborhood of
a higher rank (simple) singularity.  Following the scheme
in Malgrange [{\bf 16-18}] we use these local models to prove that
the zeros of the tau function, introduced by Jimbo,
Miwa, and Ueno in their pioneering work on ``Birkhoff''
deformations at irregular singular points [{\bf 12}], occur at
precisely those points in the deformation space at which
a certain Birkhoff-Riemann-Hilbert problem fails to have
a solution.  
\vskip.2in \par\noindent
\S 1 {\bf Introduction}
\vskip.1in\par\noindent
{\bf The Riemann-Hilbert problem and monodromy }
{\bf preserving deformations}.  Suppose that $A(x)$ is an $p\times 
p$ matrix with entries that
are rational functions of $x$ on ${\bf P}^1$. Linear differential
equations
$${{d\psi}\over {dx}}=A(x)\psi ,\eqno (1.0)$$
arise in many important applications and have been 
studied intensively for more than 100 years.  The nature of the
singularities in the coefficient matrix $A(x)$ at its poles
has a lot to do with the character of the solutions to
(1.0).  For example, in the neighborhood of a point $a$
where $A(x)$ has but a simple pole it is well known that
the equation (1.0) has a fundamental solution, $\Psi (x)$, with
pol$ $ynomially limited growth as $x\rightarrow a$ (by which we mean 
that solutions are dominated near $x=a$ by $c|x-a|^{-N}$where 
$c$ and $N$ 
are constants).  Higher order 
poles in $A(x)$ typically produce local fundamental solutions with 
more complicated exponential-polynomial growth near the
pole.  Fundamental solutions to the equation (1.0) are
usually not single valued.  If one has
a fundamental solution $\Psi (x)$ defined for $x$ in some small
ball not containing any of the poles, $\{a_1,a_2,\ldots ,a_n\}$, of $
A(x)$ then 
it is possible to analytically continue $\Psi (x)$ along paths in $
{\bf P}^1$ 
which avoid these poles. The analytic 
continuation depends only on the homotopy class of the
path, so that the resulting function, $\Psi (x)$, is
defined not on ${\bf P}^1$ but for $x$ in the simply connected 
covering space, ${\cal R}(X)$, of $X:={\bf P}^1\backslash \{a_1,a_
2,\ldots ,a_n\}$ with 
projection $\pi :{\cal R}(X)\rightarrow X$. If $x_0\in {\cal R}(X
)$ and
$\gamma$ is a closed path in $X$ with base point $\pi (x_0)$ then
$$\Psi (\gamma\cdot x_0)=\Psi (x_0)M(\gamma )^{-1},$$
where $x_0\rightarrow\gamma\cdot x_0$ is the natural action of the homotopy
class of the path $\gamma$ on $x_0\in {\cal R}(X)$, and $\gamma\rightarrow 
M(\gamma )$ is an
$n$ dimensional representation of the fundamental group
$\pi_1(X,\pi (x_0))$.
  Riemann's analysis of the solutions
of the hypergeometric equation in terms of its 
{\it monodromy representation} $M(\cdot )$ led to the general 
problem of determining if any representation of the 
fundamental group of the punctured sphere could arise 
as the monodromy representation for a linear differential 
equation (1.0).  It is clear from simple examples, that 
there will not be a unique association between a 
representation of $\pi_1$ and a differential equation (1.0)
without some further restriction on the differential
equation.  It is possibly most natural to look for
a differential equation with {\it simple poles\/} to realize a
given monodromy representation.  It is not always 
possible to solve this problem [{\bf 1}] but if one relaxes
the condition on the differential equation to admit
{\it regular singular points\/} then then it is classical that
the inverse monodromy problem always has a solution 
(a good discussion of the confusion about the status of 
the solution to this problem can also be found in [{\bf 1}]). 
A pole $a$ of $A(x)$ in (1.0) is said to be a regular
singular point if fundamental solutions to (1.0) have
at worst polynomial growth at $a$.  Because the 
fundamental solutions are defined on the simply 
connected covering and paths that wind around the point
$a$ pick up powers of the local monodromy, the precise 
notion of polynomial growth requires some restriction to
sectors in the covering space [{\bf 1}].  More important for
us, however, is a {\it deformation\/} variant of the simple pole
condition.  Suppose that for some collection of points
$\{a_1^0,a_2^0,\ldots ,a_n^0\}$ a given representation of 
$$\pi_1\left({\bf P}^1\backslash \{a_1^0,a_2^0,\ldots ,a_n^0\},a_
0\right)$$
can be realized by a differential equation (1.0) with 
{\it simple poles\/} at $a_j^0$ $j=1,2,\ldots ,n$ and a fundamental
solution $\Psi (x)$ normalized to $\Psi (a_0)=I$ at the base point
$a_0$. In this case, we write $A^0(x)$ for the matrix 
coefficient in the differential equation (1.0) which
realizes the appropriate monodromy representation.
Since $A^0(x)$ has simple poles,
$$A^0(x)=\sum_{\nu =1}^n{{A_{\nu}^0}\over {x-a_{\nu}^0}}$$
The problem, first formulated by Schlesinger [{\bf 19}], is to
ask whether it is possible to deform the coefficients
$A_{\nu}^0$ in $A^0(x)$ as functions of the pole locations $a_{\nu}$ so that
the differential equation (1.0) with coefficient matrix
$$A(x)=\sum_{\nu =1}^n{{A_{\nu}(a)}\over {x-a_{\nu}}}$$
realizes the {\it same\/} monodromy representation as as the 
differential equation with coefficient matrix (1.1) (we will
be more precise about what this means later on). 
 Note that we've written $a=(a_1,a_2,\ldots ,a_n)$ and we want
$$A_{\nu}(a^0)=A_{\nu}(a_1^0,a_2^0,\ldots ,a_n^0)=A_{\nu}^0.$$
When the point at infinity is a regular point and 
$a_0=\infty$, Schlesinger showed that if such a deformation 
exits the coefficients $A_{\nu}$ must satisfy a non-linear system of 
differential equations
$$dA_{\mu}=-\sum_{\nu\ne\mu}{{A_{\mu}A_{\nu}-A_{\nu}A_{\mu}}\over {
a_{\mu}-a_{\nu}}}d(a_{\mu}-a_{\nu})\eqno $$
now called the Schlesinger equations.  A modern 
treatment of the existence question can be found in [{\bf 17}]. 
If we write $a\in {\bf C}^n$ then as might be guessed 
from looking at the Schlesinger equations it is important 
to remove the 
points at which $a_{\mu}=a_{\nu}$ from consideration.  Let
$$\Delta_{\mu\nu}=\{a\in {\bf C}^n|a_{\nu}=a_{\mu}\},$$
and define
$$Z^n={\bf C}^n\backslash\cup_{\nu\ne\mu}\Delta_{\mu\nu}.$$
As observed by Malgrange an appropriate place to seek 
monodromy preserving deformations is the simply 
connected covering space ${\cal R}(Z^n)\rightarrow Z^n$.  Because this same
space will enter our considerations with a different 
significance later on, it will be useful at this point to 
introduce
distinctive notation for elements in ${\cal R}(Z^n)$ and their 
projection onto $Z^n$.  We will write $t\in {\cal R}(Z^n)$ and 
$a(t)\in {\cal R}(Z^n)$ for the projection of $t$ on ${\cal R}(Z^
n)\subset {\bf C}^n$. We
write $a_j(t)$ for the $j^{th}$ component of $a(t)$.

As pointed out in [{\bf 17}] and [{\bf 11}] the space ${\cal R}(Z^
n)$ has a number 
of advantages as a deformation space.  Not only is it 
simply connected but since $Z^n$ is contractible 
[{\bf 7}] the long exact sequence 
associated with the fiber bundle projection
$$a:{\cal R}(Z^n)\rightarrow Z^n$$
shows that all the higher homotopy groups of ${\cal R}(Z^n)$ are
also trivial and hence that ${\cal R}(Z^n)$ is contractible.  Since
$Z^n$ is the complement of the zero set of an an analytic
function on ${\bf C}^n$ it is a Stein space and since ${\cal R}(Z^
n)$ is an 
unramified covering of $Z^n$ it too is a Stein space [{\bf 9}].
The consequent triviality of the sheaf 
cohomology $H^1({\cal R}(Z^n),{\cal O}^{*})$ plays a role in giving a global
definition of tau functions following [{\bf 11}].  Finally, 
although it will not be the principal focus of the 
deformations we 
consider in this paper we indicate the 
crucial property used to construct the Schlesinger 
deformations 
considered in [{\bf 17}] and [{\bf 11}].  Let $Y_k$ denote the subset
of ${\bf P}^1\times {\cal R}(Z^n)$ given by
$$Y_k=\{(x,t)|x=a_k(t)\}.$$
Let 
$$Y_{\infty}=\{(\infty ,t)|t\in {\cal R}(Z^n)\},$$
and
$$Y=Y_{\infty}\cup Y_1\cup Y_2\cup\cdots\cup Y_n.\eqno (1.1)$$
Then the property, which is exploited in [{\bf 17}] to construct
the Schlesinger deformations, is that for any choice of 
$t_0\in {\cal R}(Z^n),$ the injection
$${\bf P}^1\backslash \{a_1^0,a_2^0,\ldots a_n^0,\infty \}\ni x\rightarrow 
(x,t^0)\in {\bf P}^1\times {\cal R}(Z^n)\backslash Y$$
induces an isomorphism of fundamental groups (where
$a_j^0=a_j(t^0)$ is the $j^{th}$ coordinate of the projection).  Using 
the correspondence between flat 
connections and representations of the fundamental 
group from [{\bf 6}], this allows one to prolong the original connection
from ${\bf P}^1\backslash \{a_1^0,a_2^0,\ldots ,a_n^0,\infty \}$ to $
{\bf P}^1\times {\cal R}(Z^n)\backslash Y$ in a holonomy 
preserving fashion.  The crux of the existence proof for
the Schlesinger deformations is then to extend this
connection to a neighborhood of $Y$ so that it has
{\it logarithmic poles\/} along $Y$.  In [{\bf 17}] (and by a similar 
construction in [{\bf 11}]) this is accomplished
by exhibiting local deformations in a neighborhood of each $Y_k$ 
and then proving that these can be fit together to provide a 
solution to the deformation problem if and only if a certain 
Fredholm integral equation has a solution. We will 
consider a related construction later on in this paper. 

Some further
analysis then leads to the existence of a holomorphic 
function $\tau$ defined on ${\cal R}(Z^n)$ whose 0 set is an exceptional
set for the solution of the original deformation problem. 
In this paper we wish to use similar constructions to 
examine a somewhat different class of deformations.  
Our goal will be to show that the $\tau$ function introduced 
by Jimbo, Miwa, and Ueno [{\bf 12}] in a study of such 
deformations has a similar property.  Namely,
that the zeros of these tau functions are exceptional sets
for the solution of a deformation problem.   
\vskip.1in\par\noindent
{\bf Birkhoff deformations}.  The deformations we wish to consider are associated 
with Birkhoff's generalization of the Riemann-Hilbert 
problem [{\bf 5}].  To understand what is involved it is useful to
review the analysis of solutions to linear differential equations
$${{d\psi}\over {dx}}=A(x)\psi ,$$
with a rational matrix valued coefficient $A(x)$ in a neighborhood
of an irregular singular point.  A helpful modern review of this
subject can be found in [{\bf 23}].  Suppose that $x=a$ is a 
pole of $A(x)$ of order $r+1$ then we have
$$A(x)=A_r(x-a)^{-r-1}+A_{r-1}(x-a)^{-r}+\cdots$$
The integer $r$ is called the rank of the singularity and 
for the moment
we will confine our attention to the case $r\ge 1$.  
The theory of {\it formal\/} solutions to (1.0) is much 
simplified by one further assumption which we will 
make from now on.  We require that 
\vskip.1in\par\noindent
{\bf Standing Assumption} (1.2). {\sl\ The coefficient, $A_r$, of the leading 
singularity in the Laurent expansion of $A(x)$ at $x=a$
has distinct eigenvalues.}
\vskip.1in\par\noindent
This assumption, of course, guarentees that $A_r$ can be
diagonalized by a non singular matrix $G$
$$G^{-1}A_rG=\Lambda_r$$
where $\Lambda_r$ is a diagonal matrix with distinct complex 
entries.  In such circumstances (1.0) has a unique {\it formal}
fundamental solution
$$\hat{\Psi }(x,t)=G\hat{\alpha }(x)e^{H(x)}$$
where $\hat{\alpha }(x)$ is an $n\times n$ matrix valued formal power series
$$\hat{\alpha }(x)=I+\beta_1(x-a)^1+\beta_2(x-a)^2+\cdots$$
and 
$$H(x)=\Lambda_r{{(x-a)^{-r}}\over {-r}}+\Lambda_{r-1}{{(x-a)^{-r
+1}}\over {-r+1}}\cdots +\Lambda_1{{(x-a)^{-1}}\over {-1}}+\Lambda_
0\log(x-a),$$
where all the matrices $\Lambda_k$ are diagonal with diagonal 
entries
$$\Lambda_{k,j}:=(\Lambda_k)_{jj}.$$
The 
construction of this formal solution hinges on the 
inversion of $\hbox{\rm ad}(A_r)$ acting on the off diagonal matrices.
Since the eigenvalues of $\hbox{\rm ad}(A_r)$ acting on the off diagonal
matrices are differences of distinct eigenvalues for $A_r$,
our standing assumption guarentees that this can be 
done.  It is quite typical that the series for $\hat{\alpha}$ does not
converge and this leads to some complication in making
a connection between the formal solution to (1.0) and genuine
solutions to this equation.  Before we turn to this 
matter we mention a slightly different way of looking at
this result which will make it simpler for the reader to 
connect this way of thinking with the
developments in [{\bf 17}], {\bf [11}], and [{\bf 23}].  

Now write $\partial ={{\partial}\over {\partial x}}$ and $\bar{\partial }
={{\partial}\over {\partial\bar {x}}}$ and instead of the 
differential equation (1.0) one regards the connection
$$dx\otimes (\partial -A(x))+d\bar {x}\otimes\bar{\partial },\eqno 
(1.3)$$
on the trivial bundle
$${\bf P}^1\times {\bf C}^p\rightarrow {\bf P}^1,$$
as the fundamental object. If $\{a_1,a_2,\ldots ,a_n\}$ is the set of 
poles for $A(x)$ then flat sections, $\psi$, for (1.3) defined
locally in ${\bf P}^1\backslash \{a_1,a_2,\ldots ,a_n\}$
are solutions to the differential equation (1.0).
$$ $$
Gauge transformations (e.g., multiplication by smooth 
invertible matrix valued functions of $x$) which
are holomorphic (or meromorphic) in $x$ do not change the
$d\bar {x}\otimes\bar{\partial}$ part of the connection (at least away from the
singularities) and the solution of the 
differential equation (1.0) is effectively accomplished by 
gauging $\partial -A(x)$ into diagonal form by such
a gauge transformation.  The 
formal gauge substitution
$$\psi\leftarrow G\hat{\alpha}\psi$$
can then be seen to reduce the connection (1.3) to the 
diagonal form
$$dx\otimes (\partial -h(x))+d\bar {x}\otimes\bar{\partial },$$
where $h(x):={{dH}\over {dx}}$.  A fundamental solution to the 
differential equation
$$(\partial -h(x))\psi =0,\eqno (1.4)$$
is given by
$$\Psi =e^{H(x)},$$
which is well defined modulo the possible appearance of 
a multivalued log term in $H(x)$. This accounts for the
structure of the formal fundamental solution above but
the relation between this formal solution and a genuine
fundamental solution is complicated by Stokes' 
phenomena which we will now describe.

We first adopt some notation and definitions from [{\bf 23}]. $\Sigma$ will
denote a sector with vertex at the origin, consisting of
points $re^{i\theta}$ with $r>0$ and argument $\theta\in (a,b)$ with 
$0\le a<b<2\pi$.  For $\delta >0$, we write $\Sigma_{\delta}$ for the subset of
$\Sigma$ with $r<\delta .$  If $\Sigma$ and $\Sigma'$ are two sectors we write
$\Sigma'\subset\subset\Sigma$ if the bounding rays for $\Sigma'$ are contained in $
\Sigma .$
An open set $\Omega\subset\Sigma$ is said to be asymptotic to the sector $
\Sigma$,
if for each $\Sigma'\subset\subset\Sigma$ we have $\Sigma'_{\delta}
\subset\Omega$ for all sufficiently
small $\delta$. We introduce ${\cal A}(\Sigma )$, a complex algebra of germs
of analytic functions defined on open sets asymptotic
to $\Sigma$ consisting of functions which are asymptotic to
formal meromorphic series.  The asymptotic condition
on $f\in {\cal A}(\Sigma )$ is understood to mean that there exists a
formal series 
$$\hat {f}(x)\sim\sum_{k\ge m}f_kx^k,$$
with $m>-\infty$ so that for any $\Sigma'\subset\subset\Sigma$ and any integer $
M$
one has
$$f(x)=\sum_{k\ge m}^Mf_kx^k+O(|x|^{M+1})\hbox{\rm \ as }x\rightarrow 
0\hbox{\rm \ in }\Sigma'.$$
The basic local existence result for solutions of
(1.0) near $x=a$ (and we take $a=0$ for convenience) is that
if a sector $\Sigma$ is chosen appropriately, there is a function 
(or germ) $\alpha_{\Sigma}\in {\cal A}(\Sigma )$ which is asymptotic to $
\hat{\alpha}$ with the property 
that the gauge transformation by $\alpha_{\Sigma}^{-1}$ reduces the 
differential equation (1.0) to the diagonal form (1.4) in an open
set $\Omega$ asymptotic to $\Sigma .$ To describe the non-uniqueness
for $\alpha_{\Sigma}$, upon which the Stokes' phenomena hinges, we
introduce 
$$\alpha_{\Sigma ,k}=k^{th}\hbox{\rm column of }\alpha_{\Sigma},$$
and
$$H_k(x)=k^{th}\hbox{\rm \ diagonal entry of }H(x).$$
Another way to state the existence result mentioned 
above is that the $n$ vector valued functions
$$\psi_k(x)=\alpha_{\Sigma ,k}(x)e^{H_k(x)},$$
are independent solutions to (1.0) in some open set $\Omega$ 
asymptotic to the sector $\Sigma$.  Because
the functions $\alpha_{\Sigma ,k}(x)$ are asymptotic to power series
the ``growth'' of the functions $\psi_k(x)$ as $x\rightarrow 0$ is 
controlled by the exponential factors $e^{H_k(x)}$ which have
absolute value $e^{\Re H_k(x)}$(where $\Re x=$real part of $x$). Now 
let
$$\Delta H_{jk}(x)=\Re (H_j(x)-H_k(x)).$$
The curves along which
$$\Delta H_{jk}(x)=0,\eqno (1.5)$$
play an important role in understanding the relationship 
between formal solutions and analytic solutions near the 
singularity at $x=0$. To get an idea of what such
curves look like near 0 it is enough to consider the
leading order equivalent of (1.5).  This is
$$\Re (\Lambda_{r,j}-\Lambda_{r,k})x^{-r}=0.\eqno (1.6)$$
Since each difference $\Lambda_{r,j}-\Lambda_{r,k}\ne 0$ it follows that there
are $2r$ rays emanating from 0 which satisfy (1.6) given 
by
$$\arg x={1\over r}\left(\arg(\Lambda_{r,j}-\Lambda_{r,k})+(n+{1\over 
2})\pi\right),\hbox{\rm \ for }n=0,\ldots ,2r-1.$$
These rays are called Stokes' lines. Near 
$x=0$ the family of solutions to (1.5) consists of $2r$ 
curves each asymptotic to one of the Stokes' lines 
(1.6).  The property of the Stokes' lines that will be
important for us is that any open sector which
contains one of the Stokes' lines (1.6) will contain points
with $\Delta H_{jk}<0$ and also points with  $\Delta H_{jk}>0$.

Now suppose that one crosses such a Stokes' line going
from $\Delta H_{jk}<0$ to $\Delta H_{jk}>0$.  One moves from a 
region in which $\psi_k(x)$ dominates $\psi_j(x)$ as $x\rightarrow 
0$ to a
region in which this dominance relation is reversed.
For the purpose of illustration assume that 
$\Delta H_{jk}(x)<0$ for $x$ in some truncation $\Sigma_{\delta}$ of $
\Sigma$.  Then 
for any constant $c$ one finds
$$\psi_k+c\psi_j=(\alpha_{\Sigma ,k}+c\alpha_{\Sigma ,j}e^{H_j-H_
k})e^{H_k},$$
and because $e^{H_j-H_k}$ is exponentially small in $\Sigma_{\delta}$ it follows
that
$$ $$
$$\alpha_{\Sigma ,k}+c\alpha_{\Sigma ,j}e^{H_j-H_k}\sim\hat{\alpha}_
k$$
where $\hat{\alpha}_k$ is the $k^{th}$ column of the formal series $
\hat{\alpha}$.  
Thus the less dominant solution $\psi_j$ may be freely mixed
in with $\psi_k$ without effecting the asymptotics of $\alpha_{\Sigma 
,k}$. In
this way one sees that genuine solutions with a given
exponential behavior are not uniquely determined by the
asymptotics of their ``power series component''. 
One obvious way to ``cure'' this particular non-uniqueness
would be to include a Stokes' line from the family
$\Delta H_{jk}=0$ in the sector $\Sigma$. The exchange of dominance
across the line makes it impossible to alter $\psi_k$ by adding
in multiples of $\psi_j$ without changing the asymptotics of
$\alpha_{\Sigma ,k}$ and vice versa. In fact, if one includes 
{\it exactly one Stokes' line from each of the families (1.6)}
{\it for} $j<k$ then a simple argument [{\bf 22}] shows that a
genuine fundamental solution
$$\Psi =G\alpha_{\Sigma}e^H,$$
in $\Sigma_{\delta}$ is uniquely determined by the condition that the 
asymptotic expansion of $\alpha_{\Sigma}$ is given by $\hat{\alpha}$.  What's more
there is also an existence result for such sectors which
can be proved using a variant of the usual integral 
equation technique but employing different contours for
the each of the matrix elements in the solution (see the 
references in [23]).
Following [16] we call such sectors {\it good sectors}.  It is 
not hard to see that a punctured neighborhood of 0 can
be covered by $2r$ (truncated) good sectors $\Sigma_{i,\delta}$ which we
will take to be arranged in counterclockwise order 
starting with $\Sigma_{1,\delta}$.  Because each good sector contains 
exactly one Stokes' line from each family (1.5), it follows
that the intersections $\Sigma_{i,\delta}\cap\Sigma_{i+1,\delta}$ do not contain any 
Stokes' lines.  On this overlap the two local fundamental 
solutions $\alpha_{\Sigma_i}e^H$ and $\alpha_{\Sigma_{i+1}}e^H$ of (1.0) necessarily differ by a
constant invertible $p\times p$ matrix $S_{i,i+1}$,
$$\alpha_{\Sigma_{i+1}}e^H=\alpha_{\Sigma_i}e^HS_{i,i+1}.\eqno (1
.7)$$
The matrices $S_{i,i+1}$ are called Stokes' multipliers and 
must satisfy a triangularity property which we will
now explain.  Since $\Sigma_{i,\delta}\cap\Sigma_{i+1,\delta}$ does not contain any
Stokes' lines, it follows that for each fixed choice of
$(j,k)$ the quantity $\Delta H_{jk}$ is either always positive or 
always negative in $\Sigma_{i,\delta}\cap\Sigma_{i+1,\delta}$ (at least if $
\delta$ is small
enough so that the Stokes' lines are good ``stand ins'' 
for the curves $\Delta H_{jk}=0$). Thus there is a fixed
dominance ordering 
$$\Re\Lambda_{r,i_1}x^{-r}>\Re\Lambda_{r,i_2}x^{-r}>\cdots >\Re\Lambda_{
r,i_n}x^{-r},$$
for $x\in\Sigma_{i,\delta}\cap\Sigma_{i+1,\delta}$, and some permutation $
(i_1,i_2,\ldots ,i_n)$ of
$1,2,\ldots ,n$. If we write (1.7) in matrix form relative to the
ordered basis $\{e_{i_1},e_{i_2},\ldots ,e_{i_n}\}$ then the matrix of $
S_{i,i+1}$ 
relative to this ordered basis must be lower triangular
with 1's on the diagonal in order that $S_{i,i+1}$ should only
alter $\alpha_{\Sigma_i}$ by exponentially small terms.  Another (basis 
independent) version of the same observation is that
$$e^HS_{i,i+1}e^{-H}=I+O(|x|^N),\hbox{\rm \ for }x\rightarrow 0\hbox{\rm \ in }
\Sigma_{i,\delta}\cap\Sigma_{i+1,\delta},\eqno (1.8)$$
for some $\epsilon >0$ and all positive integers $N$.  The relation 
(1.7) allows us to construct
a fundamental solution $\Psi$ to (1.0) in a neighborhood of 
$x=a$ by analytically continuing the fundamental solution $\alpha_{
\Sigma_1}e^H$
from $\Sigma_{1,\delta}$ to $\Sigma_{2,\delta}$ to $\Sigma_{3,\delta}$ and etc.  The result is
$$\Psi (x)=\left\{\eqalign{&\alpha_{\Sigma_1}(x)e^{H(x)}\hbox{\rm \ for }
x\in\Sigma_{1,\delta}\cr
&\alpha_{\Sigma_2}(x)e^{H(x)}S_{1,2}^{-1}\hbox{\rm \ for }x\in\Sigma_{
2,\delta}\cr
&\cdots\cr
&\alpha_{\Sigma_{2r}}(x)e^{H(x)}S_{1,2r}^{-1}\hbox{\rm \ for }x\in
\Sigma_{2r,\delta}\cr}
\right.\eqno (1.8)$$
where we've written 
$$S_{1,k}=S_{1,2}S_{2,3}\cdots S_{k-1,k},$$
and it is understood that the logarithmic term in $H(x)$ is
analytically continued from $\Sigma_1$ to $\Sigma_2$ to $\ldots$ to $
\Sigma_{2r}$.  If 
we write $S_{2r,1}$ 
for the Stokes' multiplier connecting $\Sigma_{2r}$ with $\Sigma_
1$ then
it is not hard to see that the analytic continuation of
$\alpha_{\Sigma_1}(x)e^{H(x)}$ around $x=a$ comes back to its original 
value multiplied on the right by,
$$e^{H(e^{2\pi i}x)-H(x)}(S_{1,2r}S_{2r,1})^{-1}=e^{2\pi i\Lambda_
0}(S_{1,2r}S_{2r,1})^{-1}.\eqno (1.9)$$
The matrix (1.9) is thus the local 
monodromy for the fundamental solution (1.8). We will
refer to the {\it exponent of formal monodromy} $\Lambda_0$, together
with the Stoke's multipliers $S_{i,i+1}$ as the generalized
monodromy data for (1.0) at $x=a$.  Roughly speaking
the deformations of (1.0) we are interested in are those 
that fix the local generalized monodromy data at each of the
singularities for (1.0) and fix the global monodromy for
(1.0) but permit the formal expansion coefficients 
$\{\Lambda_r,\Lambda_{r-1},\ldots\Lambda_1\}$ at each singularity to vary.  
The global monodromy is precisely the representation of 
the fundamental group described earlier and we will 
say a more about this later on following the 
presentation in
Malgrange [17] and Helmink [11]. Generalizations of the 
Schlesinger deformations
in which the location of the poles are varied are also 
quite interesting; however, the issues we wish to pursue
have already been treated for these deformations in [17]
and [11], and so for the moment we confine our attention to
deformations of the local exponents $\{\Lambda_r,\Lambda_{r-1},\ldots 
,\Lambda_1\}$.  Our
strategy in studying these deformations will follow 
closely the developments in [18]. In fact, in [18]
Malgrange already proves existence results for such 
deformations in the irregular singular case.  However,
we did not understand how to make use of his results
to establish the connection with the JMU tau function
that is the principal object in this paper.  Instead, we
will adapt an integral equation technique from Flaschka
and Newell [8] to produce local models for the desired
deformation at each of the poles.  These local 
deformations are then fit together by solving the same
Toeplitz integral equations that one finds in [17]. The
tau function is introduced by identifying its 
log-derivative as the connection one form for a flat 
connection on an appropriate determinant bundle. A 
computation shows that this connection one form differs
from the JMU connection one form by a regular term
and so the tau function we've introduced and the JMU 
tau functions have the same 0 set.  We show that this
0 set is precisely the exceptional set for the existence of
the deformations we are considering.
\vskip.1in\par\noindent
{\bf Local analysis and Stokes' multipliers}. It will be useful at this point to be a little more
precise about the nature of the generalized local
monodromy data that is to be ``fixed'' under the
deformations that are of interest to us. These 
deformations concern the local model for our 
connections. Suppose that one starts with a holomorphic
connection $\bar{\nabla}^0$ defined on a trivial bundle over a 
punctured neighborhood of the point $a\in {\bf P}^1$ with a
singularity of type $r$ at $a$ (note: we will put a bar over
connections defined on subsets of ${\bf P}^1$ to distinguish them 
from the connections in many variables that will soon
appear as deformations). Suppose that the connection
satisfies our standing assumption and that the
trivialization is chosen so that the leading singularity
in the one form for $\bar{\nabla}^0$ has a diagonal matrix coefficient.
Then $\bar{\nabla}^0$ is formally gauge equivalent to the diagonal form
$$d_x-d_x{\bf H}_0-\Lambda_0{{dx}\over {x-a}}\eqno (1.10)$$
where
$${\bf H}_0=\sum_{j=1}^r{{\Lambda_j^0}\over {-j}}(x-a)^{-k}\eqno 
(1.11)$$
where $\Lambda^0_j$ and $\Lambda_0$ are diagonal matrices. More precisely
there is a formal gauge transformation 
$$\hat{\alpha}^0(x)=I+\beta_1^0(x-a)+\beta_2^0(x-a)^2+\ldots$$
so that
$$\bar{\nabla}^0=\hat{\alpha}^0\cdot\left[d_x-d_x{\bf H}_0-\Lambda_
0{{dx}\over {x-a}}\right],$$
where $\hat{\alpha}\cdot [X]=\hat{\alpha }[X]\hat{\alpha}^{-1}.$
Note that it is actually the inverse of $\hat{\alpha}^0$ which reduces
$\bar{\nabla}^0$ to diagonal form.  This is just how things work out
if the relationship between $\hat{\alpha}^0$ and a fundamental solution
to $\bar{\nabla}^0$ is given along the lines explained above.

The local analytic equivalence class of $\bar{\nabla}^0$ is determined
by further data which can be specified by choosing
a covering of the punctured neighborhood of $a$ by good
sectors, $\Sigma_1,\Sigma_2,\ldots ,\Sigma_{2r}$ (to somewhat unburden the notation
we will write $\Sigma_k$ for the truncated sector $\Sigma_{k,\delta}$ when the
precise value of $\delta$ is not an issue).  For simplicity in the following
discussion we will always suppose that such a covering
is obtained by first choosing a good sector $\Sigma_1$.  The
other sectors $\Sigma_j$ are obtained from $\Sigma_1$ by rotating
counterclockwise by ${j\over {\pi}}$ radians. The Stokes' multipliers $
S_{j,j+1}$ 
which connect local fundamental solutions with
fixed asymptotics in the sectors $\Sigma_j$ and $\Sigma_{j+1}$ (with 
$\Sigma_{2r+1}=\Sigma_1$) then 
determine the local holomorphic equivalence class of $\nabla^0$.
The deformations we wish to focus on allow the local
model
$$\bar{\nabla}_{\Lambda}:=d_x-d_x{\bf H}-\Lambda_0{{dx}\over {x-a}}\eqno 
(1.12)$$
with 
$${\bf H}=\sum_{j=1}^r{{\Lambda_j}\over {-j}}(x-a)^{-j}\eqno (1.1
3)$$
to vary in the sense that the diagonal coefficient matrices
 $\Lambda_r,\Lambda_{r-1},\ldots\Lambda_1$ vary in a neighborhood of
 $\Lambda_r^0,\Lambda_{r-1}^0,\ldots\Lambda_1^0$ with $\Lambda_r$ maintaining distinct eigenvalues 
and $\Lambda_0$ held fixed.  In what follows $\Lambda_0$ will always 
denote a fixed diagonal $p\times p$ matrix.

The Stokes' multipliers are to remain fixed under our deformations.
However, since the Stokes' multipliers are significant
relative to some choice of a covering by good sectors
$\{$$\Sigma_1,\Sigma_2,\ldots ,\Sigma_{2r}\}$ we must either fix this covering for all
values of the deformation parameters (this is what is
done in [22] and [12]) or say how the
choice of covering affects the notion of ``fixed'' Stokes' 
multipliers.   We follow Malgrange in adopting
the second alternative. The choice of a good sector $\Sigma_1$ is 
nearly equivalent to the selection of a suitable collection of 
``consecutive'' Stokes' lines (one from each of the families
(1.6)).  Once one has selected such 
a collection of Stokes' lines any open sector which 
contains these Stokes' lines could serve as a good 
sector.  However, we wish to put a further limitation
on our good sectors.  No Stokes' line should lie on the
boundary of a good sector.  The reason for this is that
for a fixed connection, the Stokes' multipliers associated
with such a good sector are not stable under small 
rotations of the sector; Stokes' lines can rotate in or
out the sector and this changes the associated Stokes'
multipliers. Another way of saying this is that the
local analytic equivalence class associated with a
choice of a covering by good sectors and associated
Stokes' multipliers is not stable under small rotations
of the good sectors unless the sectors do not have
Stokes' lines on their boundary.  We will say that 
a covering $\{$$\Sigma_1,\Sigma_2,\ldots ,\Sigma_{2r}\}$ of a punctured neighborhood of $
a$ 
by good sectors is {\it stable\/} if the good sector $\Sigma_1$ has no
Stokes' lines on its boundary.
 
We now define a 
configuration space, ${\cal C}$, for our local models (1.12),
$${\cal C}:=Z^p\times {\bf C}^p\times\cdots\times {\bf C}^p,$$
where there are $r-1$ factors ${\bf C}^p$.
The first factor $Z^p$ is, of course, the configuration space
for the leading coefficient $\Lambda_r,$ and the remaining factors
${\bf C}^p$ are associated with the $\Lambda_j$, $j=r-1,\ldots ,1$.  Following
Malgrange, we now introduce a fiber bundle, 
${\cal M}\rightarrow {\cal C}$.  The fiber ${\cal M}_{\Lambda}$ over each point $
\Lambda :=(\Lambda_r,\Lambda_{r-1},\ldots ,\Lambda_1)$ in 
${\cal C}$ is the moduli space of all  holomorphic gauge 
equivalence classes of locally defined type $r$ connections 
on the trivial bundle $\{x:|x-a|<\epsilon \}\times {\bf C}^p$ (for some $
\epsilon >0$) which 
are formally equivalent to the connection (1.12-1.13).    
Or put another way, if $\nabla_1$ and $\nabla_2$ are two
type $r$ connections at $a$ and they are formally equivalent
to $\nabla_{\Lambda}$ via $\hat{\alpha}_1$ and $\hat{\alpha}$$_2$, then $
\nabla_1\simeq\nabla_2$ if $\hat{\alpha}_2^{-1}\hat{\alpha}_1$ is 
convergent in a neighborhood of $a$ (see [18]).  Next we 
want to show that 
$${\cal M}\matrix{\pi\cr
\rightarrow\cr
\cr}
{\cal C},\eqno (1.14)$$
is a fiber bundle with a natural flat connection. This 
connection will play a role in providing a global sense to
monodromy preserving deformations.  We first discuss
local trivializations for (1.14). Suppose that
$\Lambda^0\in {\cal C}$ and write $\bar{\nabla}^0$ for the connection (1.13) associated
with $\Lambda^0$. Suppose that $\{$$\Sigma_1,\Sigma_2,\ldots ,\Sigma_{
2r}\}$ is a covering
of a punctured neighborhood of $a$ by stable good sectors
for $\bar{\nabla}^0$.  Then we can find a neighborhood $U$ of $\Lambda^
0$ in ${\cal C}$ 
so that $\{$$\Sigma_1,\Sigma_2,\ldots ,\Sigma_{2r}\}$ is a stable covering at $
a$ for all 
$\bar{\nabla}_{\Lambda}$ with $\Lambda\in U$ (in fact, it is clear that $
U$ can be taken
to be of the form $U_r\times {\bf C}^p\times\cdots\times {\bf C}^
p$ where $U_r$ is a 
sufficiently small neighborhood of $\Lambda_r^0$ in $Z^p$). 
It is a result of Malgrange and Sibuya, [18,20],
that for any $\Lambda\in U$ and suitable choice of Stokes' 
multipliers $S_{j,j+1}$
there exists a type $r$ connection at $a$ with local model 
$\bar{\nabla}_{\Lambda}$ and Stokes' multipliers $S_{j,j+1}$ relative to the covering
$\{$$\Sigma_1,\Sigma_2,\ldots ,\Sigma_{2r}\}$.  This connection is not unique but its
local holomorphic gauge equivalence class is unique.
To be ``suitable'' the Stokes' multipliers must be 1 on the 
diagonal and lower triangular relative to the dominance 
ordering of $\Re (\Lambda_{r,j}x)$ for $x\in\Sigma_j\cap\Sigma_{j
+1}.$  This identifies
the fiber of ${\cal M}$ as ${\bf C}^{rp(p-1)}$, since there are $
2r$ 
intersections $\Sigma_j\cap\Sigma_{j+1}$ and ${{p(p-1)}\over 2}$ arbitrary coefficients 
for each Stokes' multiplier. The choice of a neighborhood
$U$ and a covering $\{$$\Sigma_1,\Sigma_2,\ldots ,\Sigma_{2r}\}$ which is stable 
for all $\nabla_{\Lambda}$ with $\Lambda\in U$, thus produces a trivialization
$$\pi^{-1}(U)\simeq U\times {\bf C}^{rp(p-1)}.$$
Note: The fiber ${\cal M}_{\Lambda}$ is more invariantly defined in
[18], as $H^1(S^1,\hbox{\rm St}(\bar{\nabla}_{\Lambda}))$, the first cohomology of the
Stokes' sheaf associated with the connection $\bar{\nabla}_{\Lambda}$
(see also [23]).
\vskip.2in \par\noindent
We will now show that ${\cal M}$ is a fiber bundle by 
determining that the transition maps between trivializations
are given by polynomial diffeomorphisms in the fiber.
Suppose that $U'$ is a neighborhood in ${\cal C}$ with 
$\{$$\Sigma_1',\Sigma_2',\ldots ,\Sigma_{2r}'\}$ a covering by sectors at $
a$ which is stable
for all $\bar{\nabla}_{\Lambda}$ with $\Lambda\in U'$.  Now suppose that $
\Lambda\in U\cap U'$.  
Then since $\{$$\Sigma_1,\Sigma_2,\ldots ,\Sigma_{2r}\}$ and $\{$$
\Sigma_1',\Sigma_2',\ldots ,\Sigma_{2r}'\}$ are both
good stable covering for $\bar{\nabla}_{\Lambda}$ it follows that, up to small
changes in the opening angle of $\Sigma_1'$, which do not effect
the Stokes' multipliers, $\Sigma_1'$ can be obtained from $\Sigma_
1$ by
rotating $\Sigma_1$ counterclockwise through an angle $\theta$.  To 
emphasize this we will write $\Sigma_j'=\Sigma_j^{\theta}.$  Now let $
\bar{\nabla}$ denote
a connection of type $r$ at $a$ which is formally equivalent
to $\bar{\nabla}_{\Lambda}$ (i.e., $\pi (\bar{\nabla })=\Lambda$).  Suppose $
\hat{\alpha}$ is a formal series with
$$\bar{\nabla }=\hat{\alpha}\cdot{\left[\bar\nabla_{\Lambda}\right
]}$$
in the sense of formal series at $a$.  Suppose that $\alpha_j^{\theta}
\in {\cal A}(\Sigma_j^{\theta})$ is
asymptotic to $\hat{\alpha}$ and $\bar{\nabla }=\alpha_j^{\theta}
\cdot\left[\bar\nabla_{\Lambda}\right]$ analytically in
the sector $\Sigma_j^{\theta}$.  Then the Stokes' multipliers 
$S_{j,j+1}^{\theta}$ are defined by,
$$\alpha_{j+1}^{\theta}(x)e^{H_{\Lambda}(x)}=\alpha_j^{\theta}(x)
e^{H_{\Lambda}(x)}S_{j,j+1}^{\theta},\eqno (1.15)$$
where,
$$H_{\Lambda}(x)=\sum_{j=1}^r{{\Lambda_j}\over {-j}}(x-a)^{-j}+\Lambda_
0\log(x-a).\eqno (1.16)$$
There is a slight ambiguity in the definition of the 
Stokes' multipliers in (1.15) associated with the choice
of $x\rightarrow\log(x-a)$ in (1.16).  To deal with this ambiguity
we require that the data which specifies a local 
trivialization for ${\cal M}\rightarrow {\cal C}$ includes not only the choice of
a stable good covering $\{\Sigma_1,\Sigma_2,\ldots ,\Sigma_{2r}\}$ but also a branch
of the function $x\rightarrow\log(x-a)$ in the sector $\Sigma_1$.  One may
analytically continue this choice from
$\Sigma_1$ to $\Sigma_2$ to $\Sigma_3$ and so on to fix a choice of $\log$ in (1.16)
and render (1.15) an unambiguous defining relation for
$S_{j,j+1}^{\theta}$.  In the rank $r=1$ case there are only two 
sectors $\Sigma_1$ and $\Sigma_2$ and the intersection $\Sigma_1\cap
\Sigma_2$ is 
disconnected.  In this case we suppose that the analytic
continuation from $\Sigma_1$ to $\Sigma_2$ is accomplished so that
the function is smooth on counterclockwise oriented
circles passing from $\Sigma_1$ into $\Sigma_2$.

In the special case $\theta =0$ we will simply write $\alpha_j^0=
\alpha_j$ 
and $S_{j,j+1}^0=S_{j,j+1}$. 

Now we turn to the proof that ${\cal M}\rightarrow {\cal C}$ is a fiber bundle.
As noted above our trivializations depend on a choice of
$\log(x-a)$ in $\Sigma_1^{\theta}$. However,
different choices will simply alter $S_{j,j+1}^{\theta}$ by conjugation
with powers of $\exp(2\pi i\Lambda_0)$. Since this is a linear 
transformation in the fiber, constant in the base we may 
as well suppose that this choice of $\log(x-a)$ has been 
fixed in $\Sigma_1^{\theta}$ (and $\Sigma_1$).  The Stokes' multipliers (1.15) are 
thus well defined and the transition map we wish to 
compute takes $\{S_{j,j+1}\}$ to $\{S_{j,j+1}^{\theta}\}$.
It is enough to
compute the transition in the fiber for $0<\theta <{{\pi}\over r}$ since
a larger rotation may be realized as a composition of 
rotations satisfying this condition.  Suppose now 
that $0<\theta <{{\pi}\over r}$ . 
Because the rotation $\theta$ is smaller than ${{\pi}\over r}$ it follows that
$\Sigma_j\cap\Sigma_j^{\theta}$ is not empty.  We may thus compare the local
fundamental solutions $\alpha_je^{H_{\Lambda}}$ and $\alpha_j^{\theta}
e^{H_{\Lambda}}$ on this 
intersection,
$$\alpha_je^{H_{\Lambda}}=\hbox{\rm \ $\alpha_j^{\theta}e^{H_{\Lambda}}$}
S_j(\theta ),\eqno (1.17)$$
where $S_j(\theta )$ is a constant $p\times p$ matrix.  Combining (1.16) 
and (1.17) one finds that
$$S_{j,j+1}^{\theta}=S_j(\theta )S_{j,j+1}S_{j+1}(\theta )^{-1}\eqno 
(1.18)$$
Thus to find $S_{j,j+1}^{\theta}$ in terms of the Stokes' multipliers
$\{S_{k,k+1}\}$ it will suffice to determine $S_k(\theta )$ in terms of
$\{S_{k,k+1}\}$.

By relabling the eigenvalues, $\Lambda_{r,j}\hbox{\rm \ }j=1,\ldots 
,p$ , of $\Lambda_r$ we may 
suppose that the dominance ordering in $\Sigma_j\cap\Sigma_{j+1}$ is the
``natural'' one
$$\Re (\Lambda_{r,}{}_1x)<\Re (\Lambda_{r,}{}_2x)<\cdots <\Re (\Lambda_{
r,}{}_px)$$
for $x\in\Sigma_j\cap\Sigma_{j+1}.$  It is not hard to see what happens to this
dominance ordering as one rotates $\Sigma_j\cap\Sigma_{j+1}$ counterclockwise.
As $\Sigma_j\cap\Sigma_{j+1}$ crosses a ``simple'' Stokes' line, say 
$$\Re (\Lambda_{r,}{}_1x)<\Re (\Lambda_{r,}{}_2x)=\Re (\Lambda_{r
,}{}_3x)<\Re (\Lambda_{r,}{}_3x),$$
then the dominance ordering permutes 2 and 3 leaving
the rest of the indices in sequence.  If one crosses a
Stokes' line with ``higher multiplicity'', say
$$\Re (\Lambda_{r,}{}_1x)<\Re (\Lambda_{r,}{}_2x)=\Re (\Lambda_{r
,}{}_3x)=\Re (\Lambda_{r,}{}_4x)<\Re (\Lambda_{r,}{}_5x),$$
then the (2,3,4) part of the ordering is inverted to
(4,3,2). 

For the purpose of illustration suppose that
$p=5$ and that in going from $\Sigma_j\cap\Sigma_{j+1}$ to $\Sigma_
j^{\theta}\cap\Sigma_{j+1}^{\theta}$ one passes 
through the simple Stokes' line $\Re (\Lambda_{r,}{}_1x)=\Re (\Lambda_{
r,}{}_2x)$ and the 
higher multiplicity
Stokes' line $\Re (\Lambda_{r,}{}_3x)=\Re (\Lambda_{r,}{}_4x)=\Re 
(\Lambda_{r,}{}_5x)$.  Then it is not hard
to see that the Stokes' multiplier for $\Sigma_j^{\theta}\cap\Sigma_{
j+1}^{\theta}$ must have
the following ``triangularity'',
$$S_{j,j+1}^{\theta}=\left[\matrix{1&*&0&0&0\cr
0&1&0&0&0\cr
*&*&1&*&*\cr
*&*&0&1&*\cr
*&*&0&0&1\cr}
\right]\eqno (1.19)$$
relative to the basis for ${\bf C}^5$ in which $S_{j,j+1}$ is lower 
triangular. The $*'s$ represent possibly non-zero entries.
Next consider (1.17).  Since $\alpha$$_j$ and $\alpha_j^{\theta}$ have the same
asymptotics in the sector $\Sigma_j\cap\Sigma_j^{\theta}$ it follows that $
S_j(\theta )$
must have 1's on the diagonal and
cannot have non zero off diagonal elements for any of 
the pairs associated with Stokes' lines in the intersection
$\Sigma_j\cap\Sigma_j^{\theta}$.  In the example (1.19) this means that the $
(\ell ,m)$ 
matrix elements for $S_j(\theta )$ are zero for 
$$(\ell ,m)\in \{(k,1),(k,2),(1,k),(2,k):k=3,4,5\}.$$
Since $\theta <{{\pi}\over r}$ it follows that $\Sigma_j\cap\Sigma_{
j+1}$ is contained in 
$\Sigma_j\cap\Sigma_j^{\theta}$ (provided at least one Stokes' line is crossed,
which is the only interesting case).  Thus one can also
say of $S_j(\theta )$ that it is lower triangular with respect to
the dominance ordering in $\Sigma_j\cap\Sigma_{j+1}$.  Thus for our 
example the matrix of $S_j(\theta )$ must have the form,
$$S_j(\theta )=\left[\matrix{1&0&0&0&0\cr
a&1&0&0&0\cr
0&0&1&0&0\cr
0&0&b&1&0\cr
0&0&c&d&1\cr}
\right].\eqno (1.20)$$
In a similar fashion $\Sigma_j^{\theta}\cap\Sigma_{j+1}^{\theta}$ is contained in $
\Sigma_{j+1}\cap\Sigma_{j+1}^{\theta}$
and it follows that $S_{j+1}(\theta )$ must be lower triangular 
with respect to the dominance ordering for $\Sigma_j^{\theta}\cap
\Sigma_{j+1}^{\theta}.$
In our example, if we rewrite (1.17) in the form
$S_{j,j+1}^{\theta}S_{j+1}(\theta )=S_j(\theta )S_{j,j+1}$ and make use of the lower
triangularity of the product $S_{j,j+1}^{\theta}S_{j+1}(\theta )$ with respect 
to the dominance ordering in $\Sigma_j^{\theta}\cap\Sigma_{j+1}^{
\theta}$.  We find
$$\left[\matrix{1&*&0&0&0\cr
0&1&0&0&0\cr
*&*&1&*&*\cr
*&*&0&1&*\cr
*&*&0&0&1\cr}
\right]=\left[\matrix{1&0&0&0&0\cr
a&1&0&0&0\cr
0&0&1&0&0\cr
0&0&b&1&0\cr
0&0&c&d&1\cr}
\right]\left[\matrix{1&0&0&0&0\cr
*&1&0&0&0\cr
*&*&1&0&0\cr
*&*&*&1&0\cr
*&*&*&*&1\cr}
\right]\eqno (1.21)$$
The (2,1) matrix element of $S_j(\theta )$ can be determined 
simply by equating the (2,1) matrix elements on both
sides of (1.21).  One finds
$$a=S_j(\theta )_{2,1}=-(S_{j,j+1})_{2,1}.$$
The same thing can be done for the matrix elements $b$
and $d$ for $S_j(\theta )$ on the subdiagonal.  For example
$$d=S_j(\theta )_{5,2}=-(S_{j,j+1})_{5,2}.$$
Once one knows the subdiagonal elements one can move
out to the diagonal below the subdiagonal.  Equating
(5,3) matrix elements for (1.21) one finds
$$0=c+d(S_{j,j+1})_{4,3}+(S_{j,j+1})_{5,3}.$$
From the earlier relation for $d$ one finds that $c$ is a 
polynomial function of the matrix elements of $S_{j,j+1}$.
Thus the entries of $S_j(\theta )$ are polynomials in the entries
for $S_{j,j+1}$ and it is clear that this is true quite 
generally and does not depend on anything special in our
example.  From (1.18) it then follows that $S_{j,j+1}^{\theta}$ is a 
polynomial in the entries of $S_{j,j+1}$ and $S_{j+1,j+2}$.  Since
this relation is invertible by construction it follows that
the map from $\{S_{k,k+1}\}$ to $\{S_{k,k+1}^{\theta}\}$ is a polynomial 
diffeomorphism on ${\bf C}^{rp(p-1)}$.  

We've seen that $\pi :{\cal M}\rightarrow {\cal C}$ is a fiber bundle and that there
is a natural family of trivializations for this bundle
which are related by polynomial diffeomorphisms in the
fiber that are constant in the base. 
Recall that the vertical vectors in the tangent space to
${\cal M}$ at $p$, $T_p({\cal M})$, are those killed by $d\pi_p$. A connection on
${\cal M}$, is determined by a one form $\omega$ on ${\cal M}$ whose value
$\omega_p(v)$ at a vector $v\in T_p({\cal M})$ is a vertical vector in
$T_p({\cal M})$ at $p$; $\omega$ must have the further property that
$\omega_p(v)=v$ if $v$ is a vertical vector in $T_p({\cal M}).$  Our 
trivializations single out a flat connection on ${\cal M}$.  If
$\pi^{-1}(U)$ has trivialization $U\times {\bf C}^N$ and $f_1,f_2
,\ldots f_N$ are the
natural coordinate functions on ${\bf C}^N$ then it is easy
to see that
$$\omega =\sum_{k=1}^N{{\partial}\over {\partial f_k}}df_k$$
defines a connection one form independent of the choice
of trivialization (among the special class of 
trivializations that we have been considering for ${\cal M}$
in which the fiber coordinates transform among 
themselves).  The curve $\hat{\gamma }(t)$ in ${\cal M}$ is the horizontal lift 
of $\gamma (t)=\pi\hat{\gamma }(t)$ provided that $\omega (\hat{\gamma}'
(t))=0$.  Relative to
one of our distinguished trivializations this translates
into
$$\hat{\gamma }(t)=(\gamma (t),f)$$
where $f\in {\bf C}^N$ is constant in $t$.  The curvature of this
connection is clearly 0 and locally parallel sections look
like
$$\sigma (x)=(x,f)$$
where $f$ is independent of $x$.  Now let ${\cal R}({\cal C})$ denote the
simply connected covering space of ${\cal C}.$  As was the case
for ${\cal R}(Z^n)$ above it will be convenient to introduce special
notation for projection from ${\cal R}({\cal C})$ to ${\cal C}$.  We will write
$\lambda\in {\cal R}({\cal C})$ and $\Lambda (\lambda )\in {\cal C}$ for the projection of $
\lambda$ onto ${\cal C}$.

The bundle ${\cal M}$ over
${\cal C}$ with flat connection $\omega$ pulls back to a bundle over
${\cal R}({\cal C})$ (which we will continue to denote by ${\cal M}$) with a flat
connection (which we will continue to denote by $\omega$). 
Because ${\cal R}({\cal C})$ is simply connected the bundle
$${\cal M}\rightarrow {\cal R}({\cal C})\eqno $$
has globally defined parallel sections. The existence of 
these global sections, which are the analogues of the
local notion of ``constant Stokes' multipliers'' in ${\cal C}$ is the
reason for working here in ${\cal R}({\cal C})$ instead of just locally
in ${\cal C}$ (note: the theory
of local systems explained in [23] also applies to ${\cal M}$ and
could be used to bypass the introduction of a connection
in the fiber bundle ${\cal M}$). 
\vskip.1in\par\noindent
{\bf Local models for integrable deformations}. Before we state 
our global result for ${\cal R}({\cal C})$ deformations
we introduce some notation from [18] concerning 
connections in several complex variables.  Suppose that 
$X$ is a complex analytic variety of dimension $n$, $Y$ is a
smooth hypersurface in $X$, $E$ is a rank $p$ complex vector bundle
over $X$, $\nabla$ is a holomorphic connection on $X\backslash Y$ and $
\Omega$ is 
the one form for $\nabla$ in a local frame for $E$.  
\vskip.1in\par\noindent
{\bf Definition} 1.22. {\sl For $r\ge 0$    an integer one says that 
 $\nabla$   is of type $r$ along $Y$, if in a system of local 
coordinates  
$x_1,x_2,\ldots ,x_n$   for $X$    with $Y$  locally defined by 
$x_1=0$    one has   
$$\Omega =\sum_{j=1}^nM_jdx_j,$$
where $M_1$    has a pole of order $r+1$    in $x_1$,    
$M_j$    for $j\ge 2$ 
   has a pole of order $r$    in $x_1$    and $
M_j$    is 
holomorphic in   
$x_2,x_3,\ldots x_n$    for all $j$.}
\vskip.1in\par\noindent
Suppose that 
$$M_1=\sum_{-r-1}^{\infty}M_{1,k}x_1^k.$$
Our standing assumption (1.2) for type $r\ge 1$ connections, 
translates into this situation as the assumption 
that $M_{1,-r-1}(x_2,x_3,\ldots ,x_n)$ has distinct eigenvalues. We 
will refer to such connections as {\it simple} type $r$ 
connections along $Y$.  When $r=0$ we will say that
a type 0 connection is {\it simple\/} if
$M_{1,-1}(x_2,x_3,\ldots ,x_n)$ has distinct eigenvalues which in
addition {\it do not differ by integers}.  Malgrange has 
shown how to develop the theory of Schlesinger 
deformations without this assumption, but it will be
convenient for us to insist on it so that we may work
with regular and irregular singular points in parallel.  
Propositions (1.23a) and (1.23b) below are the principal
reason that this is possible and will permit us to 
present the ``unified'' formula for $d\log\tau$ which can be
found in [12].

We recall some results of Malgrange for simple {\it integrable}
type $r$ connections. Suppose that one has a simple integrable 
type $r$ connection $\nabla$ along $Y$ and that, as above, $Y$ is 
locally defined by $x_1=0$.  Let $y=(x_2,x_3,\ldots ,x_n)$ denote the
local coordinates on $Y$.  Then one has the following local 
version of Proposition 1.3 in [18].  
\vskip.2in\par\noindent
{\bf Proposition} 1.23a  {\sl\  Suppose that $(E,\nabla )$ is a vector 
bundle
with a simple integrable connection $\nabla$ of type $r\ge 1$ along $
Y$.
There exists a local trivialization in a
neighborhood of $x=0$ so that $M_{1,-r-1}$ is diagonal in this
trivialization. Let $\Omega$ denote the connection form for 
$\nabla$ in this trivialization.
Then in a sufficiently small neighborhood $|y|<\epsilon$ 
there exists a formal power series $\hat{\alpha}$ in $x_1$  
$$\hat{\alpha }(x)=I+\beta_1(y)x_1+\beta_2(y)x_1^2+\cdots$$
with matrix coefficients $\hat{\beta}_j(y)$    which are holomorphic in 
$y$ for $|y|<\epsilon$,  with the property that  
$$\hat{\alpha}\cdot [\nabla_{\Lambda}]=d+\Omega$$
where $\hat{\alpha}\cdot [\nabla_{\Lambda}]=\hat{\alpha}\nabla_{\Lambda}
\hat{\alpha}^{-1}$ is the formal gauge 
transformation by $\hat{\alpha}$, 
$$\nabla_{\Lambda}=d-d({\bf H})-\Lambda_0{{dx_1}\over {x_1}},$$
$d$ is the exterior derivative in the $x$ 
variables, $\Lambda_0$ is a constant diagonal matrix, 
$${\bf H}=\sum_{j=1}^r\Lambda_j(y){{x_1^{-j}}\over {-j}},$$
 with $\Lambda_j(y)$    a diagonal matrix with entries that are
holomorphic functions of $y$  , and $\Lambda_r(y)$ a diagonal matrix 
with distinct entries (note: in the statement of this 
Proposition $\Lambda$ does not denote the projection 
${\cal R}({\cal C})\rightarrow {\cal C}$ described above). 
}
\vskip.1in\par\noindent
There is an analogue of this result for simple type 0
connections, which is the principal reason we work
with such connections.
\vskip.1in\par\noindent
{\bf Proposition} 1.23b. {\sl Suppose that $(E,\nabla )$ is a vector bundle
with a simple integrable type 0 connection $\nabla$ along $Y${\it .}
There exists a local trivialization near $x=0$ so that
$M_{1,-1}$ is diagonal.  Let $\Omega$ denote the one form for $\nabla$ in
this trivialization.  Then in a sufficiently small 
neighborhood of $x=0$ there exists a $\hbox{\rm GL}(p,C)$ valued holomorphic
function
$$\alpha (x)=I+\beta_1(y)x_1+\beta_2(y)x_1^2+\cdots$$
so that
$$\alpha\left(d-{{\Lambda_0}\over {x_1}}\right)\alpha^{-1}=d+\Omega 
.$$
where $\Lambda_0$ is a diagonal matrix which is independent of $x$ }.
\vskip.1in\par\noindent
Remark: The conclusion of this theorem is simply  
that the connection $\nabla$ is given by $d-{{\Lambda_0}\over {x_
1}}$ in a 
suitable local trivialization.  The analogy with 
Proposition 1.23a is not so apparent in this formulation,
however.
\vskip.1in\par\noindent
Proof (of Proposition 1.23b).  Let $\Omega =\sum_{j=1}^nM_jdx_j$ denote 
the connection one
form for $\nabla$ relative to some trivialization in a 
neighborhood of $x=0$.  Suppose $M_{1,-1}(y)$ is the residue of
$M_1$ at $x_1=0.$ Then the matrix $M_{1,-1}(0)$ has
distinct eigenvalues, so this remains true for $M_{1,-1}(y)$ 
for $x$ in a sufficiently small neighborhood of 0.  Because
its eigenvalues are distinct one may diagonalize $M_{1,-1}(y)$
by a holomorphic similarity transformation
$$Q(y)M_{1,-1}(y)Q(y)^{-1}=\Lambda_0(y),$$
where $Q(y)$ is holomorphic and $\Lambda_0(y)$ is diagonal.  Making
a gauge transformation by $Q$ one finds that the 
connection $\nabla$ becomes
$$d-{{\Lambda_0(y)}\over {x_1}}dx_1-\sum_{j=1}^nB^j(x)dx_j,$$
where the $B^j(x)$ is holomorphic in $x$.  By assumption
the eigenvalues of $\Lambda_0(0)$ do not differ by integers and
so the same is true for $\Lambda_0(y)$ if $y$ remains in a sufficiently
small neighborhood of 0.  Now let $z=x_1$ and consider the connection
$$d_z-{{\Lambda_0(y)}\over z}dz-B^1(z,y)dz,$$
depending on the parameter $y$.  Since the eigenvalues of 
$\Lambda_0(y)$ are distinct and do not differ by integers, the 
standard construction of a fundamental solution 
(which depends on the inversion of $\hbox{\rm ad}(\Lambda_0)-nI$ where
$n=1,2,\ldots$ is an integer) shows that one can 
find a gauge transformation $\beta (z,y)=I+O(z)$
which depends holomorphically on the parameters $y$ and
which transforms this connection to
$$d_z-{{\Lambda_0(y)}\over z}dz.$$
The gauge transformation of the complete connection by
this transform gives one
$$d-{{\Lambda_0(y)}\over z}dz-\sum_{j=1}^{n-1}B^{j+1}(z,y)dy_j,\eqno 
(1.23)$$
where $y_j=x_{j+1}$.  Now we express the integrability
of this connection $(d\Omega +\Omega\wedge\Omega =0)$ expanding $
B^k(z,y)$
in powers of $z$,
$$B^k(z,y)=\sum_{j=0}^{\infty}B_j^k(y)z^j.$$
Equating the coefficients of $z^{-1}$ in the integrability
condition we find that 
$${{\partial\Lambda_0(y)}\over {\partial y_j}}=\left[B_0^j,\Lambda_
0\right].$$
Since $\Lambda_0$ is diagonal the right hand side vanishes on
the diagonal and this shows that $\Lambda_0(y)=\Lambda_0=\hbox{\rm constant}
.$
The left hand side vanishes off the diagonal and since
the entries of $\Lambda_0$ are distinct this implies that $B_0^j$ must
be diagonal.  Equating the coefficients of $z^jdz\wedge dy_j$ in
the integrability condition one finds
$$nB_n^j=\left[\Lambda_0,B_n^j\right]\hbox{\rm \ for }n=1,2\ldots$$
which implies that $B_n^j=0$ for $n=1,2\ldots$ Thus the
connection $\nabla$ has the form
$$d_z-{{\Lambda_0}\over z}dz+d_y-\sum_{j=1}^{n-1}B_0^{j+1}(y)dy_j
.$$
The connection
$$d_y-\sum_{j=1}^{n-1}B_0^{j+1}(y)dy_j$$
is integrable and so a gauge transformation, $\gamma (y)$, in the
$y$ variables reduces this connection to $d_y$.  Since
$B_0^{j+1}(y)$ is diagonal this gauge transformation may be
chosen to be diagonal and also may be chosen so that
$\gamma (y)=I+O(y).$ Since $\gamma$ is diagonal it does not alter the
$dz$ part of the connection.  Composing $\beta$ and $\gamma$ we get
a gauge transformation $\alpha =I+O(z)$ which reduces
(1.23) to   
$$d-{{\Lambda_0}\over z}dz,$$
and this finishes the proof. QED
\vskip.1in\par\noindent
Although we don't require the result until the next 
section it will be convenient here to recall Theorem 2.1
from [18].  Suppose that $(E,\nabla )$ is a vector bundle with
a connection $\nabla$ that has a simple type $r$ pole along the
hypersurface $Y$ defined by $x_1=0.$ Let $y=(x_2,\ldots ,x_n)$ 
denote the coordinates along $Y$.  Suppose that in a neighborhood 
of
a point $x^0\in Y$, with coordinates $x_1=0$ and $y=0$, and 
relative to some local trivialization of 
the bundle one has a formal isomorphism,
$$\nabla =\hat{\alpha}\cdot [\nabla_{\Lambda}],$$
where $\nabla_{\Lambda}$ is the diagonal connection described in 
Proposition 1.26a and $\hat{\alpha}$ is the formal power
series described in that same proposition.  Suppose
that $\Sigma$ is a good stable sector (in the $x_1$ variable) 
for the connection $\nabla$
restricted to $y=0$.  Suppose that $\epsilon >0$ is chosen small
enough so that for all $y_1$ with $|y_1|<\epsilon$, the sector $\Sigma$ remains a good
stable sector for the restriction of $\nabla$ to $y=y_1$.  

$ $Then one has (Theorem 2.1 in [18])
\vskip.1in\par\noindent
{\bf Proposition} 1.23c. {\sl There exists a uniquely determined invertible 
holomorphic map 
$\alpha_{\Sigma}\in {\cal A}(\Sigma_{\epsilon}\times |y|<\epsilon 
)$ such that on $\Sigma_{\epsilon}\times |y|<\epsilon$ and in an appropriate 
trivialization for $E$ one has
$$\nabla =\alpha_{\Sigma}\cdot [\nabla_{\Lambda}],$$
and such that the map $\alpha_{\Sigma}$ extends $\hat{\alpha}$ in the sense that $
\alpha_{\Sigma}$ has an 
asymptotic development along $Y$ which is equal to $\hat{\alpha}$.
 }
\vskip.2in \par\noindent
Remark. The consequence of this result that is of interest for us 
is that the formal isomorphism class $\nabla_{\Lambda}$ together with 
the Stokes' multipliers determine the local holomorphic 
equivalence class of a simple, integrable type $r$ 
connection.  Two collections $\alpha_{\Sigma_k}$ and $\alpha_{\Sigma_
k}'$ associated 
with the same good stable cover $\{\Sigma_1,\ldots ,\Sigma_{2r}\}$ with the same
Stokes' multipliers and the same asymptotics $\hat{\alpha}$ clearly
differ by an invertible holomorphic map.
\vskip.2in \par\noindent

We are now 
prepared to state an existence result for a global 
version of a
``Stokes' multiplier preserving deformation'' which is
however, local in the $x_1$ variable.  The space we will
work on is $D\times {\cal R}({\cal C})$, where $D$ is the unit disk about
$x=a$ in {\bf C} and the connection whose existence we wish
to demonstrate is a simple integrable type $r$ connection
along $\{a\}\times {\cal R}({\cal C})$ which has formal reduction to 
$$\nabla_{\lambda}=d-d({\bf H})-\Lambda_0{{dx}\over {x-a}}\eqno (
1.24)$$
where $d=d_x+d_{\lambda}$ is the exterior derivative in the 
$(x,\lambda )\in D\times {\cal R}({\cal C})$ variables and 
$${\bf H}=\sum_{j=1}^r\Lambda_k(\lambda ){{(x-a)^{-j}}\over {-j}}
.\eqno (1.25)$$
Note that $\nabla_{\lambda}=e^Hde^{-H}$, is the gauge transform of the
connection $d$ by $e^H$, with $H$ given by ${\bf H}+\Lambda_0\log
(x-a)${\bf .}
However, since $H$ is singular at $x=a$ and multivalued
this is not properly a global statement but does make
sense locally.
\vskip.2in\par\noindent
{\bf Theorem} 1.26 {\sl Suppose that $\sigma$ is a parallel section for
${\cal M}\rightarrow {\cal R}({\cal C})$.  Let $D$ denote the unit disk in {\bf C} centered at
$a$.  Then there exists a holomorphic {\it integrable} connection $
\nabla$ defined
on the trivial vector bundle 
$$D\times {\cal R}({\cal C})\times {\bf C}^p\rightarrow D\times {\cal R}
({\cal C}),$$
which has a singularity of type $r\ge 1$ along the hypersurface
$Y=\{a\}\times {\cal R}({\cal C})$  such that the restriction of the connection $
\nabla$
to $D\backslash \{a\}\times {\cal R}({\cal C})$ is formally equivalent to the diagonal model 
$\nabla_{\lambda}$ defined in (1.24) and such that the the holomorphic equivalence 
class of the restriction of $\nabla$ to $(D\backslash \{a\})\times 
\{\lambda$$\}$
is given by $\sigma (\lambda )$. }
\vskip.1in\par\noindent
Proof.  We first recall a result of Malgrange and Sibuya, for
which one can also find a detailed proof in [2].  Suppose 
that $\lambda\in {\cal R}({\cal C})$ and $\sigma\in {\cal M}_{\lambda}$, where $
{\cal M}_{\lambda}$ is the fiber in ${\cal M}$ over
$\lambda .$  Then on the trivial bundle $D\times {\bf C}^p\rightarrow 
D$ there exists a
simple type $r$ connection $\bar{\nabla}$ singular at $x=a$ in $D$, which 
has formal reduction to the diagonal model
$$\bar{\nabla}_{\lambda}:=d_x-d_x{\bf H}(\lambda )-\Lambda_0{{dx}\over {
x-a}},\eqno (1.28)$$
and which has ``Stokes' multipliers'' given by $\sigma$.  In the 
version of this result that is proved in [2] the 
connection is shown to exist on a disk, $D_{\delta}$, of small radius $
\delta$.
It is not difficult to use the Birkhoff factorization 
theorem to produce a connection defined on $D\backslash \{a\}$ with the 
same properties.  Suppose then one has a connection 
$\bar{\nabla}$ defined on the trivial bundle over $D_{\delta}$ and satisfying
the conditions above.  There exists a connection $\bar{\nabla}_{e
xt}$ defined
on the trivial bundle  ${\bf C}\backslash \{a\}\times {\bf C}^p\rightarrow 
{\bf C}\backslash \{a\}$ with the same 
holonomy about $x=a$ as the connection $\bar{\nabla}$ (it is easy 
to produce such a connection with a logarithmic pole at
$x=a$).  Because the holonomy of $\bar{\nabla}$ and of $\bar{\nabla}_{
ext}$ about $x=a$
are equal it follows that there is an annulus $A$ containing
the circle, $S_{\delta /2}$, of radius $\delta /2$ on which the two 
connections are gauge equivalent.  Thus there exists a 
holomorphic map $g:A\rightarrow\hbox{\rm GL}(p,{\bf C})$ such that
$$g\cdot [\bar{\nabla }]=\bar{\nabla}_{ext}.$$
The Birkhoff theorem gives us a factorization, 
$$g=g_{\infty}^{-1}(x-a)^Ng_0$$
where $g_0$ is holomorphic and invertible in a neighborhood
of $x=a$ containing the annulus $A$, $g_{\infty}$ is holomorphic and
invertible in a neighborhood of $\infty$ which contains the 
annulus, $A,$ and $N$ is a diagonal matrix with integer entries.
The equality
$$g_0\cdot [\bar{\nabla }]=(x-a)^{-N}g_{\infty}\cdot [\bar{\nabla}_{
ext}]\eqno (1.29)$$
on the annulus $A$, shows that the connection $g_0\cdot [\bar{\nabla }
]$
extends to the punctured unit disk and since it is in the 
same local holomorphic equivalence class as $\bar{\nabla}$, we have
finished the demonstration that we may work on the 
unit disk, $D$, rather than $D_{\delta}.$

To complete the proof of Theorem 1.26 we proceed in two 
steps.  First we show that if we confine our attention 
to a sufficiently small neighborhood $U$ of $\lambda\in {\cal R}(
{\cal C})$ then
we can find a connection $\nabla_U$ defined over $D\backslash \{a
\}\times U$ which
satisfies the conclusions of Theorem 1.26, with $\sigma_U$ the
unique local flat section of ${\cal M}$ with $\sigma_U(\lambda )=
\sigma$. We will
prove this using a variant of the Flaschka-Newell 
integral equation to produce a ``perturbation'' of the 
connection $\bar{\nabla}^0$.  We defer the proof of this result to
Proposition 1.35 below.  The second step is to put 
together the ``local'' solutions $\nabla_U$ to get something 
defined on all of ${\cal R}({\cal C})$.  We will now show how to do
this.  

Let $\lambda\rightarrow\sigma (\lambda )$ denote a flat section of $
{\cal M}\rightarrow {\cal R}({\cal C})$. For each
point $\lambda\in {\cal R}({\cal C})$ there exists a neighborhood $
U(\lambda ,\sigma )$ of $\lambda$ 
in which the construction of Proposition 1.35 applies.
Let ${\cal U}$ denote a subcollection of such open neighborhoods
which is a covering for ${\cal R}({\cal C})$ and for which $U\cap 
V$ is
contractible for each pair $U,V\in {\cal U}$.  We also suppose that
each neighborhood $U\in {\cal U}$ is chosen sufficiently small so
that for $\delta >0$ small enough
there exists a sectorial covering $\{\Sigma_{1,\delta},\Sigma_{2,
\delta},\ldots ,\Sigma_{2r,\delta}\}$
of the punctured neighborhood $D_{\delta}\backslash \{a\}$ which is stable
and good for the all connections
$$d_x-d_x{\bf H}(\lambda )-\Lambda_0{{dx}\over {x-a}}$$
with $\lambda\in U$.  The construction of Proposition 1.35 shows 
that there exist holomorphic maps
$$\alpha_k^U:\Sigma_{k,\delta}\times U\rightarrow\hbox{\rm GL}(p,
{\bf C}),$$
so that 
$$\alpha_k^U\cdot [\nabla_{\lambda}]=\nabla_U\hbox{\rm \ on }\Sigma_{
k,\delta}\times U.$$
Furthermore the maps $\alpha_k^U$ are related to one another
$$\alpha_{k+1}^U=\alpha_k^US_{k,k+1}(x,\lambda )$$
where $S_{k,k+1}:\Sigma_{k,\delta}\cap\Sigma_{k+1,\delta}\times U
\rightarrow\hbox{\rm GL}(p,{\bf C})$ is a gauge 
automorphism of $\nabla_{\lambda}$ that is asymptotic to the
identity {\it to all orders\/} at $x=a$. Recall that
$$S_{k,k+1}(x,\lambda )=e^HS_{k,k+1}e^{-H},$$
where $S_{k,k+1}$ is a constant matrix and $e^H$ is well defined
once a choice of $x\rightarrow\log(x-a)$ is made for $x\in\Sigma_{
1,\delta}.$ 
Suppose that $U,V\in {\cal U}$, then for
$\lambda\in U\cap V$, the fact that $S_{k,k+1}$ does not depend on $
U$
implies that the collection of holomorphic maps
$$\alpha_k^U\left(\alpha_k^V\right)^{-1}\hbox{\rm \ for }k=1,2,\ldots 
,2r$$
defines a holomorphic map, $g_{UV}$, in a punctured neighborhood,
$D_{\delta}\backslash \{a\}\times U\cap V$, and the fact that $S_{
k,k+1}(x,\lambda )$ is asymptotic
to the identity to all orders in $(x-a)$ implies that $g_{UV}$
asymptotic to a power series 
near $x=a$ that does not depend on the sector. This 
implies that $g_{UV}$ is actually holomorphic on $D_{\delta}\times 
U\cap V$.
By construction 
$$g_{UV}\cdot [\nabla_V]=\nabla_U\hbox{\rm \ on }D_{\delta}\backslash 
\{a\}\times U\cap V.\eqno (1.30)$$
This shows that the holonomy of the connection $\nabla_U$ and
the holonomy of the connection $\nabla_V$ agree on $D\backslash \{
a\}\times U\cap V$
(since $U\cap V$ is contractible the fundamental group of
the product $D\backslash \{a\}\times U\cap V$ is determined by the first factor) 
and hence that they are holomorphically equivalent on all of
$D\backslash \{a\}\times U\cap V$.  The map $g_{UV}$ has an invertible holomorphic extension 
to all of $D\times U\cap V$ with the property that $g(0,\lambda )
=I$ for
all $\lambda\in U\cap V$.  It is not difficult to see that the gauge 
transformation $g_{UV}$ is uniquely determined by (1.30) and this
normalization.  Because of this $g_{UV}g_{VW}=g_{UW}$. 
Thus the collection $\{g_{UV}|U,V\in {\cal U}\}$ is a 
collection of transition functions for a holomorphic 
vector bundle over $D\times {\cal R}({\cal C})$.  But $D\times {\cal R}
({\cal C})$ is contractible
since both factors are, and every bundle on $D\times {\cal R}({\cal C}
)$ is thus
topologically trivial.  But $D\times {\cal R}({\cal C})$ is a Stein space.  It
is a theorem of Grauert that on a Stein space every 
topologically trivial holomorphic bundle is holomorphically
trivial [10].  Thus the bundle defined by these transition
functions is holomorphically trivial.  Thus there exists 
invertible holomorphic maps $g_U:D\times U\rightarrow\hbox{\rm GL}
(p,{\bf C})$, so that
$$g_{UV}=g_U^{-1}g_V.$$
Equation (1.30) becomes
$$g_V\cdot [\nabla_V]=g_U\cdot [\nabla_U],$$
and we see that $g_U\cdot [\nabla_U]$ defines a global connection
on $D\backslash \{a\}\times {\cal R}({\cal C})$ which satisfies the conditions of Theorem 
1.26. QED
\vskip.1in \par\noindent
Now we turn to the proof of the perturbation result 
used in the proof of the preceeding theorem.
Suppose that $\bar{\nabla}^0$ is a simple type $r$ 
connection on the trivial bundle
$$D\times {\bf C}^p\rightarrow D,$$
with singularity at $x=a$ in $D$.  For the purpose of a 
technical result later on it will be convenient to choose 
a special trivialization in which to consider the 
connection $\bar{\nabla}^0$.  Choose $r<1$ and let $A$ denote the annulus
$$A=\{x:r<|x-a|<1\}\subset D.$$
Choose a point $p\in A$ and let $M$ denote the $p\times p$ 
invertible matrix which gives the holonomy of the 
connection $\bar{\nabla}^0$ (with respect to the initial trivialization)
along a counterclockwise oriented circle of radius $|p-a|$ 
about $a$.  Let 
$$m:={1\over {2\pi i}}\log M,$$
for some choice of a logarithm for $M$.  The connection
$$\bar{\nabla}^{\infty}:=d_x-{m\over {x-a}},$$
defined on the trivial bundle ${\bf P}^1\times {\bf C}^p$ has regular singular
points at 0 and $\infty$ and its restriction to $A$ has the same holonomy 
representation as $\bar{\nabla}^0$.  Hence there exists an invertible
holomorphic map $g:A\rightarrow\hbox{\rm GL}(p,{\bf C})$ so that
$$g\cdot\bar{\nabla}^0=\bar{\nabla}^{\infty}\hbox{\rm \ on }A.$$
Let $g=g_{\infty}^{-1}(x-a)^{-N}g_0$ denote the Birkhoff factorization 
of $g$, with $g_0$ holomorphic and invertible in $D$, $g_{\infty}$ 
holomorphic and invertible in $\{x:|x-a|>r\}\cup \{\infty \}$, and
$N$ a diagonal matrix with integer entries.  Thus
$$g_0\cdot\bar{\nabla}^0=(x-a)^Ng_{\infty}\cdot\bar{\nabla}^{\infty}
,$$
and we see that by adjusting the trivialization on 
$D\times {\bf C}^p$ by $g_0$ we may suppose that our connection
$\bar{\nabla}^0$ extends to a connection on the trivial bundle ${\bf P}^
1\times {\bf C}^p\rightarrow {\bf P}^1$ 
with a regular singular point at $\infty$ (the resulting 
connection may not be of simple type at $\infty$, however).  
In what follows we suppose that we are looking at $\bar{\nabla}^0$ in just
such a trivialization.
 
Now suppose that the leading
singularity in the connection one form for $\bar{\nabla}^0$ is diagonal
and that for the formal power series
$$\hat{\alpha}^0=I+\beta_1^0(x-a)+\beta_2^0(x-a)^2+\cdots$$
we have
$$\bar{\nabla}^0=\hat{\alpha}^0\cdot\left[d_x-d_x{\bf H}(\lambda^
0)-\Lambda_0{{dx}\over {x-a}}\right],\eqno (1.31)$$
where {\bf H} is given by (1.25) and $\lambda^0$ is some fixed element
in ${\cal R}({\cal C})$ covering $\Lambda (\lambda^0)$. Note that the projection $
\Lambda^0=\Lambda (\lambda^0)$ 
is determined by $\bar{\nabla}^0$ through (1.31).  Let 
$\{\Sigma_{1,\delta},\Sigma_{2,\delta},\ldots ,\Sigma_{2r,\delta}
\}$ denote a stable good covering of a 
punctured neighborhood of $x=a$ for the diagonal 
connection $\bar{\nabla}$$_{\lambda^0}$(see 1.28).  Let $\alpha_{
k^{}}^0=\alpha_{\Sigma_k}^0\in {\cal A}(\Sigma_k)$ be a holomorphic
function whose asymptotics are given by $\hat{\alpha}$ and for which
one has the analytical relation,
$$\bar{\nabla}^0=\alpha_k^0\cdot\left[d_x-d_x{\bf H}(\lambda^0)-\Lambda_
0{{dx}\over {x-a}}\right],\eqno (1.32)$$
in the sector $\Sigma_{k,\delta}$.  Finally suppose that on $\Sigma_{
k,\delta}\cap\Sigma_{k+1,\delta}$ 
we have
$$\alpha_{k+1}^0=\alpha_k^0S_{k,k+1}(x,\lambda^0)\eqno (1.33)$$
where
$$S_{k,k+1}(x,\lambda )=e^{H(\lambda )}S_{k,k+1}^0e^{-H(\lambda )}
,\eqno (1.34)$$
Here $H$ is given by (1.16) and a choice of $\log(x-a)$ is fixed 
for $x\in\Sigma_{1,\delta}$ to make $H$ well defined.  The Stokes' 
multiplier $S_{k,k+1}^0$ is independent of $x$ and $\lambda^0$.  Let $
\Omega^0$ 
denote the connection form for $\bar{\nabla}^0$ and let $pr$ denote the
natural projection
$$D\times {\cal R}({\cal C})\rightarrow D.$$
Define
$$\nabla^0=d+pr^{*}\Omega^0,$$
with $d=d_x+d_{\lambda}$, so that $\nabla^0$ defines a connection on the
trivial bundle,
$$D\times {\cal R}({\cal C})\times {\bf C}^p\rightarrow D\times {\cal R}
({\cal C}).$$

The following proposition demonstrates the existence of local 
``Birkhoff deformations'' of the connection $\bar{\nabla}^0$ in the space 
$D\times {\cal R}({\cal C}).$
\vskip.2in \par\noindent
{\bf Proposition} 1.35. {\sl For a sufficiently small neighborhood 
$U$ of $\lambda^0$ there exists a simple integrable type $r$ connection  
$\nabla_U$ defined on the trivial bundle
$$D\times U\times {\bf C}^p\rightarrow D\times U$$
so that 
\vskip.1in\par\noindent\ (i) $\nabla_U$ is formally reducible to the diagonal form 
(1.28), 
$$\nabla_U=\hat{\alpha}\cdot [\nabla_{\lambda}]$$
where
$$\hat{\alpha }=I+\beta_1(\lambda )(x-a)+\beta_2(\lambda )(x-a)^2
+\cdots$$
is a formal power series with {\it holomorphic\/} matrix valued
coefficients $\beta_k(\lambda )$.  The sectors $\Sigma_k\times U$ are
good stable sectors for $\nabla_U$; there exist holomorphic maps
$$\alpha_k\in {\cal A}(\Sigma_k,U),$$
with asymptotics given by $\hat{\alpha}$ so that on open sets 
asymptotic to $\Sigma_k\times U$,
$$\nabla_U=\alpha_k\cdot [\nabla_{\lambda}].\eqno (1.39)$$
\vskip.1in\par\noindent(ii) $\nabla_U$ is a ``Birkhoff deformation'' of $
\bar{\nabla}^0$ in that the
restriction of $\nabla_U$ to $D\backslash \{a\}\times \{\lambda^0
\}$ is equivalent to $\bar{\nabla}^0$
and
$$\alpha_{k+1}(x,\lambda )=\alpha_k(x,\lambda )S_{k,k+1}(x,\lambda 
)\eqno (1.40)$$
on $\Sigma_{k,\delta}\cap\Sigma_{k+1,\delta}\times U$ (see 1.34) 
\vskip.1in\par\noindent(iii) On the punctured neighborhood 
$D\backslash \{a\}\times U,$ the connections $\nabla_U$ and $\nabla^
0$ are gauge equivalent
$$\nabla_U=\Phi\cdot\left[\nabla^0\right]\eqno (1.41)$$
by a gauge transformation $x\rightarrow\Phi (x,\lambda )$ which is holomorphic 
in the exterior of the disk $D$ and asymptotic to $I$ as
$x\rightarrow\infty$.
\vskip.1in\par\noindent
(iv) If $\bar{\nabla}^0$ extends to a connection on the trivial bundle
${\bf P}^1\times {\bf C}^p\rightarrow {\bf P}^1$ with a regular singular point at infinity, then
$$d_{\lambda}\hbox{\rm Res}_{x=a}\hbox{\rm Tr}\left\{\hat\alpha^{
-1}d_x\hat\alpha d_{\lambda}H(\lambda )\right\}=0.\eqno (1.42)$$
 } 
\vskip.1in\par\noindent
Proof.  To construct $\nabla_U$ through equation (1.39) it will 
suffice to construct functions $\alpha_k$ satisfying (1.40) with
$\alpha_k(x,\lambda^0)=\alpha_k^0(x)$.  Our strategy will be to look for 
solutions 
$$\alpha_k(x,\lambda )=\varphi_k(x,\lambda )\alpha_k^0(x)$$
with  $\varphi_k:\Sigma_{k,\delta}\times U\rightarrow\hbox{\rm GL}
(p,{\bf C})$ a holomorphic map with
appropriate asymptotics. Condition (1.40) for $\{\alpha_k\}$ 
translates into 
$$\varphi_{k+1}(x,\lambda )=\varphi_k(x,\lambda )(I+\Delta s_{k,k
+1}(x,\lambda )),\eqno (1.43)$$
for $(x,\lambda )$ in $\Sigma_{k,\delta}\cap\Sigma_{k+1,\delta}\times 
U$, where
$$I+\Delta s_{k,k+1}(x,\lambda )=\alpha_k^0(x)S_{k,k+1}(x,\lambda 
)S_{k,k+1}(x,\lambda^0)^{-1}\alpha_k^0(x)^{-1}.$$
The important property of $\Delta s_{k,k+1}$ for us is that for 
any fixed $\delta >0$ one can make $\Delta s_{k,k+1}$ as close to  
zero as one likes by choosing $\lambda\in U$, with $U$ a sufficiently
small neighborhood of $\lambda^0$.

The condition (1.40) (or its translation (1.43)) does not 
determine $\alpha_k$ uniquely.  We will impose a further 
condition on $\alpha_k$ which will uniquely determine it.  The
extra condition is that the ``fundamental solutions'' $\Psi (x,\lambda 
)$ and
$\Psi^0(x)$ associated to $\{\alpha_k\}$ and $\{\alpha^0_k\}$ by (1.8) differ on the
circle of radius $\delta$ by a map which is holomorphic in the
exterior of the circle of radius $\delta$ about $x=a$ and asymptotic 
to the 
identity at $\infty .$ More precisely $\Psi (x,\lambda )\Psi^0(x)^{
-1}$ should be 
holomorphic in $x$ outside the circle of radius $\delta$ and
$$\Psi (x,\lambda )\Psi^0(x)^{-1}=I+O(x^{-1}).$$
Note that $\Phi (x,\lambda )=\Psi (x,\lambda )\Psi^0(x)^{-1}$ will be the gauge 
transformation in (iii) of Proposition 1.35.
We will now translate the conditions we've outlined into
an integral equation for the functions $\varphi_k$.  It will be 
useful to begin by describing the pieces that make up the  
integration contours we will use.  Let $\sigma_{k,k+1}$ denote an 
oriented ray segment that lies in the intersection 
$\Sigma_{k,\delta}\cap\Sigma_{k+1,\delta}$, joins the point $a$ to the circle of 
radius $\delta$ about $a$, and separates the Stokes' lines in the
sector $\Sigma_{k,\delta}$ from the Stokes' lines in the sector $
\Sigma_{k+1,\delta}$ 
for all $\lambda\in U$ (Stokes' lines are associated with the 
function $\Lambda_r(\lambda )$). It is always possible to do this if $
\delta$ is
small enough and $U$ is chosen to be a sufficiently small
neighborhood of $\lambda^0$.  Let $\gamma_k$ denote the counterclockwise
oriented segment of the circle of radius $\delta$ about $a$ which
joins the end point of $\sigma_{k-1,k}$ to the end point of $\sigma_{
k,k+1}$.
Finally let $\sigma_k$ denote the open wedge that is bounded by
$\sigma_{k-1,k}$, $\gamma_k$, and $\sigma_{k,k+1}$.  Then clearly $
\sigma_k\subset\Sigma_{k,\delta}$ and the
oriented boundary of $\sigma_k$ is given by
$$\partial\sigma_k=\sigma_{k-1,k}+\gamma_k-\sigma_{k,k+1}.$$
If we now compare $\Psi$ and $\Psi^0$ defined by (1.8) on the 
circle of radius $\delta$ about $a$ we find
$$\Psi (x,\lambda )\Psi^0(x)^{-1}=\varphi_k(x,\lambda )(I+\Delta 
m_k(x,\lambda ))\hbox{\rm \ for }x\in\gamma_k\eqno (1.44)$$
where $k=1,2,\ldots ,2r$ with 
$$I+\Delta m_k(x,\lambda ):=\alpha_k^0(x)e^{H(x,\lambda )-H(x,\lambda^
0)}\alpha_k^0(x)^{-1}.\eqno (1.45)$$
We've written (1.45) in the special form $I+\Delta m_k$ to 
emphasize the fact that for fixed $\delta$ the right hand side
of (1.45) can be made as close to the identity as one
pleases by choosing $\lambda\in U$, with $U$ a sufficiently small
neighborhood of $\lambda^0$.  Next we will obtain a system of
integral equations for $\varphi_k$ following Flaschka and Newell 
[8].  Suppose that $y\in\sigma_1$, then by Cauchy's theorem
$$\eqalign{\varphi_1(y,\lambda )&=\int_{\partial\sigma_1}{{\varphi_
1(x,\lambda )}\over {x-y}}{{dx}\over {2\pi i}}\cr
&=\int_{\sigma_{0,1}}{{\varphi_1(x,\lambda )}\over {x-y}}{{dx}\over {
2\pi i}}+\int_{\gamma_1}{{\varphi_1(x,\lambda )}\over {x-y}}{{dx}\over {
2\pi i}}-\int_{\sigma_{1,2}}{{\varphi_1(x,\lambda )}\over {x-y}}{{
dx}\over {2\pi i}}\cr}
\eqno (1.46)$$
On $\sigma_{1,2}$ we can use (1.43) to write
$$\eqalign{\varphi_1(x,\lambda )&=\varphi_1(x,\lambda )-\varphi_2
(x,\lambda )+\varphi_2(x,\lambda )\cr
&=-\varphi_1\Delta s_{1,2}(x,\lambda )+\varphi_2(x,\lambda ),\cr}
\eqno (1.47)$$
where for brevity we've written $\varphi_1\Delta s_{1,2}(x,\lambda 
)$ for
$\varphi_1(x,\lambda )\Delta s_{1,2}(x,\lambda )$ (we will use this notation without 
further comment in what follows).
Substituting this expression in the $\sigma_{1,2}$ integral in (1.46) 
one finds
$$\eqalign{\varphi_1(y,\lambda )=\int_{\sigma_{0,1}}{{\varphi_1(x
,\lambda )}\over {x-y}}{{dx}\over {2\pi i}}+\int_{\gamma_1}{{\varphi_
1(x,\lambda )}\over {x-y}}{{dx}\over {2\pi i}}+\int_{\sigma_{1,2}}{{
\varphi_1\Delta s_{1,2}(x,\lambda )}\over {x-y}}{{dx}\over {2\pi 
i}}\cr
-\int_{\sigma_{1,2}}{{\varphi_2(x,\lambda )}\over {x-y}}{{dx}\over {
2\pi i}}.\cr}
\eqno (1.48)$$
Since $\varphi_2(x,\lambda )$ is holomorphic in $\sigma_2$ and $y
\in\sigma_1$ which is 
outside of $\sigma_2$ it follows that
$$-\int_{\sigma_{1,2}}{{\varphi_2(x,\lambda )}\over {x-y}}{{dx}\over {
2\pi i}}=\int_{\gamma_2}{{\varphi_2(x,\lambda )}\over {x-y}}{{dx}\over {
2\pi i}}-\int_{\sigma_{2,3}}{{\varphi_2(x,\lambda )}\over {x-y}}{{
dx}\over {2\pi i}}.\eqno (1.49)$$
We substitute (1.49) for the last integral to appear in
(1.48). In the expression that results we observe that the
integral,
$$\int_{\sigma_{2,3}}{{\varphi_2(x,\lambda )}\over {x-y}}{{dx}\over {
2\pi i}}=-\int_{\sigma_{2,3}}{{\varphi_2\Delta s_{2,3}(x,\lambda 
)}\over {x-y}}{{dx}\over {2\pi i}}+\int_{\sigma_{2,3}}{{\varphi_3
(x,\lambda )}\over {x-y}}{{dx}\over {2\pi i}},$$
and one may continue this last integral to $\gamma_3$ and $\sigma_{
3,4}$ 
by Cauchy's theorem as above.  Proceeding all the way 
around the circle in this fashion one finds
$$\varphi_1(y,\lambda )-\sum_{k=1}^{2r}\left\{\int_{\gamma_k}{{\varphi_
k(x,\lambda )}\over {x-y}}{{dx}\over {2\pi i}}+\int_{\sigma_{k,k+
1}}{{\varphi_k\Delta s_{k,k+1}(x,\lambda )}\over {x-y}}{{dx}\over {
2\pi i}}\right\}=0\eqno (1.50)$$
Now we formulate the condition that the right hand side 
of (1.44) should be holomorphic in the exterior of the 
circle of radius $\delta$ about $a$ and asymptotic to the identity
at $\infty$.  This can be expressed as
$$\sum_{k=1}^{2r}\int_{\gamma_k}{{\varphi_k(x,\lambda )(I+\Delta 
m_k(x,\lambda ))}\over {x-y}}{{dx}\over {2\pi i}}=I.\eqno (1.51)$$
Adding this result to the preceeding equation one finds
$$\varphi_1(y,\lambda )+K_1\varphi (y,\lambda )=I\eqno (1.52)$$
 where
$$K_1\varphi (y,\lambda )=\sum_{k=1}^{2r}\left\{\int_{\gamma_k}{{
\varphi_k\Delta m_k(x,\lambda )}\over {x-y}}{{dx}\over {2\pi i}}-
\int_{\sigma_{k,k+1}}{{\varphi_k\Delta s_{k,k+1}(x,\lambda )}\over {
x-y}}{{dx}\over {2\pi i}}\right\}\eqno (1.53)$$
and $\varphi =(\varphi_1,\varphi_2,\ldots ,\varphi_{2r}).$
We ``derived'' (1.52) with $y$ chosen to be in the interior 
of $\sigma_1.$ However, we now choose to think of (1.52) as an
integral equation for the restriction of $\varphi_1$ to $\partial
\sigma_1$. In this
case the integral operator $K_1$ defined in (1.52) is 
understood to involve non tangential limits for $y$ on 
$\partial\sigma_1$ from the interior of $\sigma_1$.  There is was nothing 
special about $\varphi_1$ in the 
arguments above and so we find that the vector $\varphi$
satisfies the system of integral equations
$$\varphi_k(y,\lambda )+K_k\varphi (y,\lambda )=I\eqno (1.54)$$
where the integral operators $K_k$ are defined by the same
formula as $K_1$ but the $y$ variable which occurs in (1.53)
takes values in $\partial\sigma_k$ (with the integral operator defined
by non-tangential limits from the interior). It is well
known that non-tangential limits for the Cauchy kernel 
$(x-y)^{-1}$ determine a bounded operator on $L^2(\partial\sigma_
k)$ in case
both $x$ and $y$ are in $\partial\sigma_k$. It is not hard to see from this
that the integral operator 
$$K\varphi =(K_1\varphi ,K_2\varphi ,\ldots ,K_{2r}\varphi )$$
is a bounded operator on
$${\cal H}=L^2(\partial\sigma_1)\oplus L^2(\partial\sigma_2)\oplus
\cdots\oplus L^2(\partial\sigma_{2r}).$$
Furthermore it is clear that because $\Delta m_k(x,\lambda )$ and
$\Delta s_{k,k+1}(x,\lambda )$ can be made uniformly small by choosing
$\lambda$ close enough to $\lambda^0$, the system of integral equations
(1.54) has a unique solution in ${\cal H}$ provided the 
neighborhood $U$ is small enough.

Next we wish to show that the solution of (1.54) 
satisfies (1.43).  Suppose then that $\varphi$ is a solution
to (1.54) in ${\cal H}$. One calculates that for $y\in\sigma_{j,j
+1}$, 

$$\eqalign{\varphi_{j+1}(y,\lambda )-\varphi_j(y,\lambda )=K_j\varphi 
(y,\lambda )-K_{j+1}\varphi (y,\lambda )\cr
=\int_{\sigma_{j,j+1}}\left\{{{\varphi_j\Delta s_{j,j+1}(x,\lambda 
)}\over {x-y(\sigma_j)}}-{{\varphi_j\Delta s_{j,j+1}(x,\lambda )}\over {
x-y(\sigma_{j+1})}}\right\}{{dx}\over {2\pi i}}\cr}
\eqno (1.55)$$
where we've written $y(\sigma_k)$ for the boundary value on 
$\sigma_{j,j+1}$ taken from the interior of $\sigma_k$ for $k=j$ and 
$k=j+1$.  Since
$${1\over {2\pi i}}\left\{{1\over {x-y(\sigma_j)}}-{1\over {x-y(\sigma_{
j+1})}}\right\}=\delta (x-y)\hbox{\rm \ for }y\in\sigma_{j,j+1}$$
it follows from (1.55) that
$$\varphi_{j+1}(y,\lambda )-\varphi_j(y,\lambda )=\varphi_j(y,\lambda 
)\Delta s_{j,j+1}(y,\lambda )\hbox{\rm \ for }y\in\sigma_{j,j+1}.\eqno 
(1.56)$$
This is a ``boundary value'' version of (1.43), which we 
will now extend to a sectorial neighborhood of $\sigma_{j,j+1}.$ As
a simple consequence of satisfying the integral equation
(1.54) we know that $\varphi_j(y,\lambda )$ for $y\in\sigma_{j,j+
1}$ is the 
boundary value of a holomorphic function $\varphi_j(y,\lambda )$ for 
$y\in\sigma_j.$ Also $\varphi_{j+1}(y,\lambda )$ for $y\in\sigma_{
j,j+1}$ is the boundary 
value of a holomorphic function $\varphi_{j+1}(y,\lambda )$ for $
y\in\sigma_{j+1}.$
Solving (1.56) for $\varphi_{j+1}(y,\lambda )$ we see this function has an
analytic continuation into a sector containing $\sigma_{j,j+1}$, 
since equation (1.43) shows that the function 
$I+\Delta s_{j,j+1}(y,\lambda )$ (and its inverse) has an analytic 
continuation into such a sector.   One can use
Morera's theorem to show that the function obtained by
gluing together $L^2$ boundary values along $\sigma_{j,j+1}$ is actually 
holomorphic in a neighborhood of $\sigma_{j,j+1}.$ The same 
argument works for $\varphi_j(y,\lambda )$ and equation (1.56) extends 
to a sectorial neighborhood of $\sigma_{j,j+1}$.  This is (1.43). 

The iterative solution to (1.54) produces an analytic 
function of $\lambda$ with values in ${\cal H}.$ The asymptotics for
$\varphi_j$ are obtained by substituting 
$${1\over {x-y}}={1\over x}\sum_{n=0}^N\left({y\over x}\right)^n+
y^{N+1}{{x^{-N-1}}\over {x-y}}$$
into the integral equation (1.54) and noting that the
functions
 
$$\varphi_j(x,\lambda )x^{-k}\Delta s_{j,j+1}(x,\lambda )$$
 are integrable in
$x$ on $\sigma_{j,j+1}$ for all integers $k\ge 0$ with integrals that
are  analytic in $\lambda .$ 

This finishes the proof of (i), (ii).  To establish (iii) note 
that we've shown that each solution $\varphi_k$ to (1.54) extends 
to a holomorphic function in a sector containing $\sigma_k$, with
controlled asymptotic behavior as $x\rightarrow a.$ This is all that 
is needed to establish the analogue of (1.50) for $\varphi_k$. 
Subtracting this from (1.54) one obtains the analogue of 
(1.51) for $\varphi_k$.  As noted above this is an expression of 
the holomorphic character of the gauge transformation
$\Phi (x,\lambda ):=\Psi (x,\lambda )\Psi^0(x)^{-1}$ in the exterior of $
D$ and this 
finishes the proof of (iii). 

To establish (iv) (which will play an important role in
a tau function calculation in section 3) write
$$\Omega =\Omega_x+\Omega_{\lambda},$$
for the one form associated with $\nabla_U$. Here $\Omega_x$ is the
$dx$ term in the one form and $\Omega_{\lambda}$ is a sum
$$\Omega_{\lambda}=\sum_k\Omega_{\lambda ,k}d\lambda_k.$$
The $d\lambda$ component of the formal equivalence 
$\nabla_U=\hat{\alpha}\cdot\left[\nabla_{\lambda}\right]$ is
$$-d\hat{\alpha}\hat{\alpha}^{-1}-\hat{\alpha }dH\hat{\alpha}^{-1}
=\Omega_{\lambda},$$
or
$$d\hat{\alpha }=-\hat{\alpha }dH-\Omega_{\lambda}\hat{\alpha }.\eqno 
(1.57)$$
For simplicity we write $d=d_{\lambda}$ and calculate,
$$\eqalign{d\hbox{\rm Res}_{x=a}\hbox{\rm Tr}\left(\hat\alpha^{-1}
d_x\hat\alpha dH\right)&\cr
=\hbox{\rm Res}_{x=a}\hbox{\rm Tr}&\left(-\hat\alpha^{-1}d\hat\alpha
\hat\alpha^{-1}d_x\hat\alpha dH-\hat\alpha^{-1}d_xd\hat\alpha dH\right
)\cr
=\hbox{\rm Res}_{x=a}\hbox{\rm Tr}&\left(d_x(dH)dH+\hat\alpha^{-1}
d_x\Omega_{\lambda}\hat\alpha dH\right),\cr}
$$
where to get from the second to the the third line 
we substituted (1.57) for $d\hat{\alpha}$ did an obvious 
cancellation in the result and made use of the fact that
$dHdH=0$ since $dH$ is diagonal.  But
$$\hbox{\rm Res}_{x=a}\hbox{\rm Tr}\left(d_x(dH)dH\right)=0,$$
since the Laurent series for $d_x(dH)dH$ begins with terms
$C(x-a)^{-3}$ and
so we find
$$d\hbox{\rm Res}_{x=a}\hbox{\rm Tr}\left(\hat\alpha^{-1}d_x\hat\alpha 
dH\right)=\hbox{\rm Res}_{x=a}\hbox{\rm Tr}\left(d_x\Omega_{\lambda}
\hat\alpha dH\hat\alpha^{-1}\right).\eqno (1.58)$$
A straightforward calculation now shows that

$$\eqalign{\hbox{\rm Res}_{x=a}\hbox{\rm Tr}&\left(d_x(\hat\alpha 
dH\hat\alpha^{-1})\hat\alpha dH\hat\alpha^{-1}\right)\cr
&=\hbox{\rm Res}_{x=a}\hbox{\rm Tr}\left(d_x(dH)dH\right)=0\cr}
\eqno (1.59)$$
In (1.59) we replace $\hat{\alpha }dH\hat{\alpha}^{-1}$ by $-d\hat{
\alpha}\hat{\alpha}^{-1}-\Omega_{\lambda}$ from (1.57),
and find 
$$\hbox{\rm Res}_{x=a}\hbox{\rm Tr}\left(d_x(\Omega_{\lambda})\Omega_{
\lambda}\right)+2\hbox{\rm Res}_{x=a}\hbox{\rm Tr}\left(d_x(\Omega_{
\lambda})d\hat\alpha\hat\alpha^{-1}\right)=0,\eqno (1.60)$$
where we made use of the fact that $d_x(d\hat{\alpha}\hat{\alpha}^{
-1})d\hat{\alpha}\hat{\alpha}^{-1}$ is
``regular'' at $x=a$ and so has 0 residue, and that
$$\hbox{\rm Res}_{x=a}d_x\hbox{\rm Tr}\left(\Omega_{\lambda}d\hat
\alpha\hat\alpha^{-1}\right)=0.$$
Now substitute $-d\hat{\alpha}\hat{\alpha}^{-1}-\Omega_{\lambda}$ for $
\hat{\alpha }dH\hat{\alpha}^{-1}$ in (1.58) and 
make use of (1.60) to get,
$$d\hbox{\rm Res}_{x=a}\hbox{\rm Tr}\left(\hat\alpha^{-1}d_x\hat\alpha 
dH\right)=-{1\over 2}\hbox{\rm Res}_{x=a}\hbox{\rm Tr}\left(d_x(\Omega_{
\lambda})\Omega_{\lambda}\right).\eqno (1.61)$$
Now for the first time we use the fact that $\bar{\nabla}^0$ is 
looked at in a trivialization in which it extends to a 
connection on ${\bf P}^1$ with an additional regular singularity at
$\infty$.  We see from this and (iii) that,
$$\Omega_{\lambda}=-d\Phi\Phi^{-1},\eqno (1.60)$$
is holomorphic in a neighborhood of $\infty$.  Since 
$\hbox{\rm Tr}(d_x(\Omega_{\lambda})\Omega_{\lambda})$ is meromorphic on $
{\bf P}^1$ with a single pole at 
$x=a$, it follows that the residue at this pole must be 0.
With (1.61) this finishes the proof of (iv) (incidentally, the argument
here follows the argument in [12]  used to show that 
the Jimbo, Miwa, Ueno expression for $d\log\tau$ is closed). 
QED.
\vskip.2in \par\noindent
\S 2 {\bf The Vector Bundle Deformation of Malgrange }
\vskip.2in \par\noindent
{\bf Representations of the fundamental group and flat }
{\bf connections}. In this section
we are interested in constructing an integrable deformation of a
connection on a bundle over $\bf P^1$ which is monodromy
preserving and which respects the local character of the connection
near its singular points.

The principal tool in the construction of this deformation
away from the singular set is a correspondence between
representations of the fundamental group and vector 
bundles with flat connections.  More precisely, suppose
that $X$ is a connected complex manifold with base point $x^0$.  
Suppose that $E\rightarrow X$ is a complex vector bundle with 
a flat holomorphic connection $\nabla$.  Suppose that 
$\gamma :[0,1]\rightarrow X$ is a piecewise smooth curve in $X$ and 
let ${\cal P}_{\nabla}(\gamma )$ denote parallel translation with respect to
$\nabla$ along $\gamma$.  Then
$${\cal P}_{\nabla}(\gamma ):E_{\gamma (0)}\rightarrow E_{\gamma 
(1)},$$
is a linear isomorphism between the fibers of $E$ at
the endpoints $\gamma (0)$ and $\gamma (1)$.  Now suppose that $\gamma$ is
a piecewise smooth closed loop based at $x^0$ and let $g=[\gamma 
]$ denote the 
homotopy class of $\gamma$.  Then 
$$\rho (g):={\cal P}_{\nabla}(\gamma )^{-1},\eqno (2.1)$$
defines a representation of $\pi_1(X,x^0)$ on $E_{x^0}.$ The
right hand side depends only on the homotopy class
of $\gamma$ because the curvature of $\nabla$ is zero.  The 
equivalence class of the
representation $\rho$ actually determines the pair $(E,\nabla )$ up
to isomorphism.  Before we turn to the main theorem of this section we
digress to sketch a construction that
takes one from $\rho$ to $(E,\nabla )$.  Let $\pi :{\cal R}(X)\rightarrow 
X$ denote the simply
connected covering space of $X$ and suppose that $(E,\nabla )$ is a 
vector bundle with flat connection over $X$ as above.
The pull back bundle $\pi^{*}(E)\rightarrow {\cal R}(X)$ is necessarily trivial
since the base ${\cal R}(X)$ is simply connected.  The natural
projection $\tilde{\pi }:\pi^{*}(E)\rightarrow E$ is a local diffeomorphism and since
$d\tilde{\pi}_{\tilde {p}}$ is an isomorphism of tangent spaces which maps
the vertical vectors in $T_{\tilde {p}}(\pi^{*}(E))$ bijectively onto the
vertical vectors in $T_p(E)$ we may use $d\tilde{\pi}_{\tilde {p}}$ to lift the
horizontal subspace in $T_p(E)$ that comes from $\nabla$ to
a horizontal subspace in $T_{\tilde {p}}(\pi^{*}(E))$. We write $
\pi^{*}(\nabla )$ for
the resulting connection on $\pi^{*}(E),$ and note that since
the pull back connection $\pi^{*}(\nabla )$ is related to $\nabla$ by a local
diffeomorphism it is also a flat connection.  We may
thus produce
a trivialization for $\pi^{*}(E)$ consisting of flat sections
for $\pi^{*}(\nabla )$.  Let $\tilde {x}^0$ denote a base point in $
{\cal R}(X)$ such that
$\pi (\tilde {x}^0)=x^0.$ For $u\in\pi^{*}(E)_{\tilde {x}^0}$ let $
\tilde {x}\rightarrow {\cal P}(\tilde {x},u)\in\pi^{*}(E)_{\tilde {
x}}$ denote the
parallel section of $\pi^{*}(E)$ which agrees with $u$ at $\tilde {
x}^0$.
Writing $E_0:=\pi^{*}(E)_{\tilde {x}^0}$(which is naturally isomorphic to
$E_{x^0}$) we have
$${\cal R}(X)\times E_0\ni (x,u)\rightarrow {\cal P}(x,u)\in\pi^{
*}(E),$$
is an isomorphism between the bundle $\pi^{*}(E)$ and
the trivial bundle ${\cal R}(X)\times E_0$.  If we compose this
map ${\cal P}$ with the vector bundle projection  
$$\tilde{\pi }:\pi^{*}(E)\rightarrow E,$$
one obtains a vector bundle map
$$\tilde{\pi }{\cal P}:R(X)\times E_0\rightarrow E,\eqno (2.2)$$
which covers the projection $\pi :{\cal R}(X)\rightarrow X$. 

The representation 
$\rho$ from (2.1) determines a left action  of $\pi_1(X,x^0)$ on $
R(X)\times E_0$ 
given by 
$$g\cdot (\tilde {x},u)=(g\cdot\tilde {x},\rho (g)u),\eqno (2.3)$$
where $g\cdot\tilde {x}$ is just the usual action of $\pi_1$ on the simply
connected covering space ${\cal R}(X)$.  We will show that the 
quotient bundle $\pi_1(X,x^0)\backslash {\cal R}(X)\times E_0\rightarrow
\pi_1(X,x^0)\backslash {\cal R}(X)$ is 
isomorphic to $E\rightarrow X$ as a vector bundle with connection.

We begin by showing 
that the map $\tilde{\pi }{\cal P}$ is equivariant for this action of 
$\pi_1(X,x^0)$.  To see this suppose that $\tilde {x}=[\chi ]$ where $
\chi$ is
a smooth curve joining $x^0$ to $x$ in $X$.  Let $\tilde {x}^0$ denote the
class of the constant path starting and ending at $x^0$.
Write $\tilde{\chi}$ for the lift of $\chi$ into ${\cal R}(X)$ with initial point
$\tilde {x}^0$.  Let $g=[\gamma ]$ where $\gamma$ is a smooth closed path in $
X$ 
based at $x^0$. Then one finds
$$\eqalign{\tilde{\pi }{\cal P}(g\cdot\tilde {x},\rho (g)u)=&\tilde{
\pi }{\cal P}_{\pi^{*}(\nabla )}(\tilde{\gamma\chi} )\rho (g)u\cr
=&{\cal P}_{\nabla}(\gamma\chi )\rho (g)u\cr
=&{\cal P}_{\nabla}(\chi ){\cal P}_{\nabla}(\gamma )\rho (g)u\cr
=&{\cal P}_{\nabla}(\chi )u=\tilde{\pi }{\cal P}(\tilde {x},u),\cr}
$$
which shows the equivariance of the map $\tilde{\pi }{\cal P}$.  In the 
third and last equality we used the fact that 
$\tilde{\pi }{\cal P}_{\pi^{*}(\nabla )}(\tilde{\gamma })u={\cal P}_{
\nabla}(\gamma )u,$ where $\tilde{\gamma}$ is a lift of $\gamma$ and 
$u\in E_{\gamma (0)}.$ This is obvious if the curve $\gamma$ stays in a
neighborhood $U$ in $X$ which is evenly covered by the
projection on $X$ from ${\cal R}(X)$; the general
result follows from the fact that parallel translation is
an anti-homomorphism under homotopy composition.

Since the 
vector bundle action (2.3) covers the standard left
action of $\pi_1(X,x^0)$ on ${\cal R}(X)$
and since $\pi_1(X,x^0)$$\backslash {\cal R}(X)\simeq X$ one finds that
$$\pi_1(X,x^0)\backslash {\cal R}(X)\times E_0:={\cal R}(X)\times_{
\rho}E_0,$$
is a vector bundle over $X$ isomorphic to $E$ through the 
map induced by (2.2).  Since the construction of the
bundle
$${\cal R}(X)\times_{\rho}E_0\rightarrow X\eqno (2.4)$$
depends only on the representation $\rho$ it follows that
this representation determines the bundle $E\rightarrow X$ up to
isomorphism.  In fact the bundle (2.4) has a naturally defined
flat connection $\nabla_{\rho}$ so that the map induced by (2.2)
determines an isomorphism,
$$({\cal R}(X)\times_{\rho}E_0,\nabla_{\rho})\simeq (E,\nabla ).\eqno 
(2.5)$$
In order to define $\nabla_{\rho}$ we introduce a family of 
trivializations for the vector bundle,
$$\pi_{\rho}:{\cal R}(X)\times_{\rho}E_0\rightarrow X.\eqno (2.6)$$
Let ${\cal F}$ denote a covering
of $X$ by open sets, with the following properties,
\par\noindent
(1) If $U\in {\cal F}$ then $U$ is evenly covered by 
$\pi :{\cal R}(X)\rightarrow X.$ That is there exist disjoint sets $
U_{\alpha}\subset X$ such
that $\pi^{-1}(U)=\cup_{\alpha}U_{\alpha}$ and $\pi :U_{\alpha}\rightarrow 
U$ is a diffeomorphism for
each $\alpha$.
\par\noindent
(2) If $U,V\in {\cal F}$ and $U\cap V\ne\phi$ then $U\cup V$ is evenly covered
by $\pi$.
\par\noindent
The existence of such a covering is an easy consequence 
of the fact that $X$ has a metric topology.  Now suppose
that $U\in {\cal F}$ and $U_{\alpha}\subset {\cal R}(X)$ is such that $
\pi :U_{\alpha}\rightarrow U$ is a 
diffeomorphism.  We define a trivialization, $\phi (U_{\alpha})$ of $
\pi_{\rho}^{-1}(U)$
by sending each equivalence class, 
$$[(\tilde {x},u)]:=\{(y,v):(\tilde {y},v)=g\cdot (\tilde {x},u)\hbox{\rm \ for }
g\in\pi_1(X,x^0)\}$$
in $\pi_{\rho}^{-1}(U)$ into the 
unique representative of the form $(\tilde {x},u)$ with $\tilde {
x}\in U_{\alpha}$ and
$u\in E_0$.  Then 
$$\phi (U_{\alpha})[(\tilde {x},u)]:=(\pi (\tilde {x}),u)=(x,u)\in 
U\times E_0.$$
Now suppose that $U,V\in {\cal F}$ and $\pi :U_{\alpha}\rightarrow 
U$ and 
$\pi :V_{\beta}\rightarrow V$ are diffeomorphisms (note that $U=V$ is a 
possibility).  Then if $U\cap V\ne\phi$ it follows that there
exists a unique $g_{\alpha\beta}\in\pi_1(X,x^0)$ so that $U_{\alpha}
\cap g_{\alpha\beta}V_{\beta}\ne\phi .$ 
The existence of such a $g_{\alpha\beta}$ is trivial and  
uniqueness is equivalent to the assertion that if 
$U_{\alpha}\cap V_{\beta}\ne\phi$ and $U_{\alpha}\cap gV_{\beta}\ne
\phi$ for some $g\in\pi_1(X,x^0)$
then $g=1$.  Suppose then that $U_{\alpha}\cap V_{\beta}\ne\phi$ and 
$U_{\alpha}\cap gV_{\beta}\ne\phi$.  Then since $U\cup V$ is evenly covered
we have $(U_{\alpha}\cup V_{\beta})\cap g(U_{\alpha}\cup V_{\beta}
)=\phi$ if $g\ne 1$.  But 
evidently,
$$U_{\alpha}\cap gV_{\beta}\subset (U_{\alpha}\cup V_{\beta})\cap 
g(U_{\alpha}\cup V_{\beta})=\phi ,$$
so $U_{\alpha}\cap gV_{\beta}=\phi$ if $g\ne 1$.  Uniqueness follows.  One
may now easily compute
$$\phi (U_{\alpha})\phi (V_{\beta})^{-1}(x,u)=(x,\rho (g_{\alpha\beta}
)u).\eqno (2.7)$$
Since these transition functions are constant in the base
variables there is a globally defined flat connection
$\nabla_{\rho}$ on ${\cal R}(X)\times_{\rho}E_0$ which is obtained by gluing together
the exterior derivative in the base variables defined in
each of the trivializations $\phi (U_{\alpha}).$ In these trivializations
it is not hard to check the isomorphism of connections 
(2.5).  

To finish this account of the reconstruction of a 
bundle with a flat connection and prescribed holonomy
it is still necessary to check that $({\cal R}(X)\times_{\rho}E_0
,\nabla_{\rho})$ has
holonomy at $x^0$ given by $\rho$.  Suppose that $\gamma$ is a 
piecewise smooth loop in $X$ based at $x^0$ with
$g=[\gamma ]\in\pi_1(X,x^0)$.  Let $\gamma^{-1}({\cal F})$ be the open covering of
the interval [0,1] by the inverse image of sets from
${\cal F}$ under $\gamma .$ Let $\delta >0$ be a Lebesque number for this 
covering and suppose that $0=t_0<t_1\cdots <t_n=1$ is a 
partition of [0,1] with $t_{j+1}-t_j<\delta$ for all $j$.  Then
each curve segment $\{\gamma (t):t\in [t_j,t_{j+1}]\}$ lies inside some
 $U_j\in {\cal F}$.  For each $j$ one can find $U_{\alpha_j}\subset 
{\cal R}(X)$ so that
$\pi :U_{\alpha_j}\rightarrow U_j$ is a diffeomorphism and one can also arrange
that $U_{\alpha_{j+1}}\cap U_{\alpha_j}\ne\phi$ for $j=0,\ldots n
-1$.  Then parallel 
translation of a vector $u$ in $E_0$ along $\gamma$ is constant in
the trivializations $\phi (U_{\alpha_j})$ for $j=0,\ldots ,n-1$.  To compute
the holonomy one must only compute what the vector $u$
in the trivialization $\phi (U_{\alpha_{n-1}})$ looks like in the 
trivialization $\phi (U_{\alpha_0})$.  However, it is clear from the 
construction that $gU_{\alpha_0}\cap U_{\alpha_{n-1}}\ne\phi .$  Thus the parallel
transport of $u\in E_0$ along $\gamma$ gives $\rho (g)^{-1}u\in E_
0$. This 
finishes our sketch of the reconstruction of a vector 
bundle with flat connection (up to equivalence) from 
its holonomy representation.  We now begin  to explain the
setting for Theorem 2.9 below.
\vskip.2in \par\noindent
{\bf The vector bundle deformation}. Suppose that $\{a_1^0,a_2^0,
\ldots ,a_n^0\}$ is a collection of $n$ distinct
points in ${\bf C}$. In this section we will construct a global 
deformation for a connection $\bar{\nabla}^0$ defined on the trivial bundle
$E^0:={\bf P}^1\times {\bf C}^p,$ with a  
simple type $r_j$ singularity at $a_j^0$ and a regular point or 
simple type $r_{\infty}$ singularity at $\infty$. 
For simplicity in stating results, when $\bar{\nabla}^0$ has a regular
point at $\infty$ we
put $r_{\infty}=-1$ and say that $\bar{\nabla}^0$ has
a simple type -1 singularity. 

Roughly speaking the deformation we consider will preserve 
the local type of
the singularities for $\bar{\nabla}^0$ and also the local and
global monodromy data for the connection $\bar{\nabla}^0$.  We now 
make this more precise.

By relabling the points if necessary we may suppose 
that $r_j\ge 1$ for $j=1,2,\ldots ,m$ and $r_j=0$ for 
$j=m+1,m+2,\ldots ,n$. It could happen that $m=0$ or $m=n$. 
The point $\infty$ is somewhat special
in this context since it does not contribute to
the space of pole deformations, ${\cal R}(Z^n),$ which we defined
in section 1.  We will mention  special 
considerations concerning $\infty$ when we encounter them.

For $j=1,\ldots ,m$ let    
$${\cal C}_j:=Z^p\times {\bf C}^p\times\cdots\times {\bf C}^p\hbox{\rm \ with }
r_j-1\hbox{\rm \ factors }{\bf C}^p,$$
denote the local configuration space at $a_j^0$, as described in 
section 1. Recall that
each point in ${\cal C}_j$ corresponds to a formal equivalence class
for a simple type $r_j$ connection at $a_j^0$. Write ${\cal C}_{\infty}$ for the
corresponding configuration space at $\infty$, defined if
$r_{\infty}\ge 1.$

Write $\Lambda_j^0\in {\cal C}_j$ 
for the data associated to $\bar{\nabla}^0$ at $a_j^0$, for $j=1,
\ldots m.$
 Recall that ${\cal R}(X)$ is just the simply connected cover of 
$X$, and define

$${\cal D}:={\cal R}(Z^n)\times\prod_{j=1}^m{\cal R}({\cal C}_j)\times 
{\cal R}({\cal C}_{\infty}),\eqno (2.8)$$
where the product is just the Cartesian product, and the
final factor ${\cal R}({\cal C}_{\infty})$ only appears if $r_{\infty}
\ge 1$.
  
The space ${\cal D}$ will serve as our ``deformation space''.  In the
first factor, $Z^n$, is the space of pole locations and in each
of the subsequent factors, ${\cal C}_j$, is the space of local formal 
equivalence classes at $a_j^0.$ We must pass to the simply
connected cover in ${\cal D}$ to guarentee global existence for 
the sort of deformation we are about to describe.

Let ${\cal M}_j\rightarrow {\cal C}_j$ denote the fiber bundle over $
{\cal C}_j$ whose fiber 
over $\Lambda\in {\cal C}_j$ is the holomorphic equivalence class of connections
formally equivalent to the diagonal model (1.12) associated to
the base point $\Lambda$ (defined in section 1). As in section 1 we also 
write ${\cal M}_j\rightarrow {\cal R}({\cal C}_j)$ for the
pull back of ${\cal M}_j$ under the projection ${\cal R}({\cal C}_
j)\rightarrow {\cal C}_j$.  Let 
$\sigma_j^0\in {\cal M}_j$ denote the point in the fiber associated to the
class of $\bar{\nabla}^0$ in a neighborhood of $x=a_j^0$ $(\sigma_{
\infty}^0$ is also 
defined if $r_{\infty}\ge 1$).  Let $\lambda\rightarrow\sigma_j(\lambda 
)$ denote the unique flat
section of ${\cal M}_j\rightarrow {\cal R}({\cal C}_j)$ with $\sigma_
j(\Lambda_j^0)=\sigma_j^0$ ($\sigma_{\infty}(\lambda )$ is defined
in a similar fashion if $r_{\infty}\ge 1$)

Recall
that $Y_k$ is the subset of points $(x,t)\in {\bf P}^1\times {\cal R}
(Z^n)$ with
$x=a_k(t)$, and for $j=1,2,\ldots ,n$  define
$${\cal Y}_k=Y_k\times\prod_{j=1}^m{\cal R}({\cal C}_j)\times {\cal R}
({\cal C}_{\infty})\subset {\bf P}^1\times {\cal D},$$
where the factor ${\cal R}({\cal C}_{\infty})$ is present only if $
r_{\infty}\ge 1.$
Let $t^0\in {\cal R}(Z^n)$ denote a point in the covering space of $
Z^n$ such 
that $a_j(t^0)=a_j^0$, let $\lambda_j^0\in {\cal R}({\cal C}_j)$ denote a point in the 
covering space of ${\cal C}_j$ such that $\Lambda_j^0=\Lambda_j(\lambda_
j^0)$ (where $\Lambda_j$ is 
the projection from ${\cal R}({\cal C}_j)$ to ${\cal C}_j$) and write 
$${\bf P}^1\ni x\rightarrow i(x):=(x,t^0,\lambda^0)\in {\bf P}^1\times 
{\cal D},$$
where 
$$\lambda^0=(\lambda_1^0,\ldots ,\lambda_m^0,\lambda_{\infty}^0),$$
and as above $\lambda_{\infty}^0$ only occurs when $r_{\infty}\ge 
1$.

The following theorem is due to Malgrange  ([18] 
theorem 3.1),
\vskip.1in\par\noindent
{\bf Theorem} 2.9 {\sl There exists a rank $p$ holomorphic vector
bundle $E\rightarrow {\bf P}^1\times {\cal D}$ and an integrable connection $
\nabla$ on $E$ with 
a simple type $r_j$ singularity along ${\cal Y}_j$ for $j=1,\ldots 
,n$ and
a simple type $r_{\infty}$ singularity along ${\cal Y}_{\infty}$ such
that the restriction of $(E,\nabla )$ to ${\bf P}^1\times \{(t^0,
\lambda^0)\}$ is equivalent
to $(E^0,\bar{\nabla}^0)$ (that is, $i^{*}(E,\nabla )\simeq (E^0,
\bar{\nabla}^0)$). Furthermore for
$j=1,\ldots m$ the restriction of $(E,\nabla )$ to ${\bf P}^1\times 
\{(t,\lambda )\}$ is formally
equivalent to the model connection $\bar{\nabla}_{\Lambda_j(\lambda_
j)}$ (1.12) near $x=a_j(t)$
and is in the holomorphic equivalence class $\sigma_j(\lambda_j)\in 
{\cal M}_j$.
}
\vskip.1in\par\noindent
Proof. We will prove this result as Malgrange does by 
first constructing the deformation in the complement of
$${\cal Y}=\cup_{j=1}^n{\cal Y}_j\cup {\cal Y}_{\infty}\subset {\bf P}^
1\times {\cal D},$$
and then extending the connection $\bar{\nabla}^0$ to a tubular 
neighborhood ${\cal T}({\cal Y}_j)$ $({\cal T}({\cal Y}_{\infty})
)$ of each singular set ${\cal Y}_j$ $({\cal Y}_{\infty})$ so 
that it has the right local characteristics.  In particular one
finds that the two constructions must be holomorphically 
equivalent on ${\cal T}({\cal Y}_k)\backslash {\cal Y}_k$ and this equivalence allows one 
to define a bundle over all ${\bf P}^1\times {\cal D}$ together with a 
connection that has the right global and local properties.

An important result for the construction of the 
deformation on ${\bf P}^1\times {\cal D}\backslash {\cal Y}$ is the following 
observation of Malgrange.
Choose some point $x^0\in {\bf C}$ so that $x^0\ne a_j(t^0)$ for all
$j=1,\ldots ,n$.  Define $p^0=(x^0,t^0,\lambda^0)$.  Then the map
$${\bf P}^1\backslash \{a_1^0,a_2^0,\ldots ,a_n^0,\infty \}\ni x\rightarrow 
(x,t^0,\lambda^0)\in {\bf P}^1\times {\cal D}\backslash {\cal Y},\eqno 
(2.10)$$
induces an isomorphism of fundamental groups
$$\pi_1\left({\bf P}^1\backslash \{a_1^0,a_2^0,\ldots ,a_n^0,\infty 
\},x^0\right)\simeq\pi_1\left({\bf P}^1\times {\cal D}\backslash 
{\cal Y},p^0\right).\eqno (2.11)$$
This is explained in both [17] and [11] where the 
deformation space does not include the factors
${\cal R}({\cal C}_j).$ However, the product of these factors is simply connected
so it does not influence the result.  The holonomy of
the connection $\bar{\nabla}^0$ at the base point $x^0$ determines a
representation, $\rho ,$ of $\pi_1\left({\bf P}^1\backslash \{a_1^
0,a_2^0,\ldots ,a_n^0,\infty \},x^0\right)$ on $GL(p,{\bf C})$.
The isomorphism (2.11) and the representation $\rho$ 
determines a $GL(p,{\bf C})$ representation of $\pi_1({\bf P}^1\times 
{\cal D}\backslash {\cal Y},p^0)$ 
which we continue to denote by $\rho$.  Associated with this
representation is a vector bundle $E_{\rho}:=({\bf P}^1\times {\cal D}
\backslash {\cal Y})\times_{\rho}{\bf C}^p$ 
with connection $\nabla_{\rho}$ whose holonomy representation is 
given by $\rho$.

Next we turn to the construction of the local 
deformations.  Suppose that $a=(a_1,a_2,\ldots ,a_n)\in Z^n$ and
let $\delta (a)$ denote the minimum of the distances, 
$\{$$|a_i-a_j|,|a_i^0-a_j^0|\}_{i\ne j}$ (the reason for insisting that
$\delta (a)\le\min\{|a_i^0-a_j^0|\}_{i\ne j}$ will appear below).
Let $D_j(a)$ denote the disk of radius $\delta (a)/3$ about
the point $a_j$.  Let $D_{\infty}(a)$ denote the open complement of the
closed disk of radius $\delta (a)+\max_i\{|a_i|,|a_i^0|\}.$  It is clear by 
construction that the disks $\{D_{\infty}(a),D_j(a),j=1,\ldots ,n
\}$ are pairwise
disjoint for each $a\in Z^n$.  Define a tubular neighborhood, 
${\cal T}({\cal Y}_j),$ of ${\cal Y}_j$ by
$${\cal T}({\cal Y}_j)=\{(x,t,\lambda )\in {\bf P}^1\times {\cal D}
|x\in D_j(a(t))\},$$
and a tubular neighborhood of ${\cal Y}_{\infty}$ by
$${\cal T}({\cal Y}_{\infty}):=\{(x,t,\lambda )\in {\bf P}^1\times 
{\cal D}|x\in D_{\infty}(a(t))\}.$$
Then the neighborhoods $\{{\cal T}({\cal Y}_{\infty}),{\cal T}({\cal Y}_
j),j=1,\ldots n\}$ are 
pairwise disjoint.  Following the scheme that can
be found in Malgrange [17] we define a connection on
the trivial bundle ${\cal T}({\cal Y}_j)\times {\bf C}^p\rightarrow 
{\cal T}({\cal Y}_j)$ by lifting the connection
on the trivial bundle over $D_j(a^0)\times {\cal R}({\cal C}_j)$ that one
obtains from theorem 1.26 above.  Recall that 
$\lambda\rightarrow\sigma_j(\lambda )$ is the unique flat section of $
{\cal M}_j\rightarrow {\cal R}({\cal C}_j)$
with $\sigma_j(\Lambda_j^0)=\sigma_j^0.$  For $j=1,\ldots ,m$ let $
\nabla_j$ denote the integrable
connection on the trivial bundle
$$D_j(a^0)\times {\cal R}({\cal C}_j)\times {\bf C}^p\rightarrow 
D_j(a^0)\times {\cal R}({\cal C}_j)$$
with a simple type $r_j$ singularity along $\{a_j^0\}\times {\cal R}
({\cal C}_j)$,
whose existence is guarenteed by theorem 1.26.  This
connection naturally extends to a connection on the
trivial bundle
$$D_j(a^0)\times {\cal R}({\cal C})\times {\bf C}^p\rightarrow D_
j(a^0)\times {\cal R}({\cal C}),$$
by pulling back the connection one form under the 
natural projection
$$D_j(a_0)\times {\cal R}({\cal C})\rightarrow D_j(a_0)\times {\cal R}
({\cal C}_j),$$
where we've written
$${\cal R}({\cal C}):=\prod_{j=1}^m{\cal R}({\cal C}_j)\times {\cal R}
({\cal C}_{\infty}).$$
 We continue to 
denote this connection by $\nabla_j$.  Now define a map
$$pr_j:{\cal T}({\cal Y}_j)\rightarrow D_j(a^0)\times {\cal R}({\cal C}
),\eqno (2.12)$$
by $pr_j(x,t,\lambda )=(x-a_j(t)+a_j^0,\lambda )$ (the extra condition that
$\delta (a)\le\min\{|a_i^0-a_j^0$$|\}_{i\ne j}$ now guarentees that 
$x-a_j(t)+a_j^0\in D_j(a^0)$ for $x\in D_j(a^0)$.
 Let $\Omega_j(x,\lambda_j)$ denote
the one form for $\nabla_j$ (and we've written $\Omega_j(x,\lambda_
j)$ to 
emphasize the fact that $\Omega_j$ only depends on the variables $
(x,\lambda_j)$),
$$\nabla_j=d+\Omega_j(x,\lambda_j),\eqno (2.13)$$
and define a connection (which we again call $\nabla_j$!) on the
trivial bundle $E_j:={\cal T}({\cal Y}_j)\times {\bf C}^p\rightarrow 
{\cal T}({\cal Y}_j)$ by
$$\nabla_j:=d+pr_j^{*}(\Omega_j),\eqno (2.14)$$
where $d$ denotes the exterior derivative on ${\cal T}({\cal Y}_j
)$ (acting
on ${\bf C}^p$ valued functions). It is easy to check that 
connection (2.14) is integrable and has a simple type $r_j$ 
singularity along ${\cal Y}_j$ as a consequence of (2.13) being 
integrable with a simple type $r_j$ singularity along
$\{a_j^0\}\times {\cal R}({\cal C}).$  Now we wish to determine the holonomy
for $\nabla_j$ on ${\cal T}({\cal Y}_j)\backslash {\cal Y}_j$.  The map 
$${\cal T}({\cal Y}_j)\backslash {\cal Y}_j\ni (x,t,\lambda )\rightarrow 
t\in {\cal R}(Z^n),\eqno (2.15)$$
is surjective with fiber over $t$ given by 
$D_j(a(t))\backslash \{a(t)\}\times {\cal R}({\cal C})$ (which is homeomorphic to 
$D_j(a^0)\backslash \{a_j^0\}\times {\cal R}({\cal C})$).
The first part of the homotopy exact sequence for this
fiber bundle reads
$$\rightarrow\pi_2({\cal R}(Z^n))\rightarrow\pi_1({\cal T}({\cal Y}_
j)\backslash {\cal Y}_j)\rightarrow\pi_1(D_j(a^0)\backslash \{a_j^
0\})\rightarrow\pi_1({\cal R}(Z^n))\rightarrow 0.$$
where we substituted $\pi_1(D_j(a^0)\backslash \{a_j^0\})$ for 
$\pi_1(D_j(a^0)\backslash \{a_j^0\}\times {\cal R}({\cal C}))$.
Thus we have
$$\pi_1({\cal T}({\cal Y}_j)\backslash {\cal Y}_j)\simeq\pi_1(D_j
(a^0)\backslash \{a_j^0)).\eqno (2.16)$$
Let 
$$\left({\cal T}({\cal Y}_j)\backslash {\cal Y}_j\right)_{t=t^0}=
\{(x,t^0,\lambda )\in {\cal T}({\cal Y}_j)\backslash Y_j\}\simeq 
D_j(a^0)\backslash \{a_j^0\}\times {\cal R}({\cal C})$$
Then the restriction of $pr_j$,
$$pr_j:\left({\cal T}({\cal Y}_j)\backslash {\cal Y}_j\right)_{t=
t^0}\rightarrow D_j(a^0)\backslash \{a_j^0\}\times {\cal R}({\cal C}
),\eqno (2.17)$$
is essentially the identity.  Since (2.16) shows that 
representatives of all the homotopy classes of curves
in ${\cal T}({\cal Y}_j)\backslash {\cal Y}_j$ can be found among the loops that stay in
the section $\left({\cal T}({\cal Y}_j)\backslash {\cal Y}_j\right
)_{t=t^0}$ it follows that the holonomy
of the connection $\nabla_j$ can be computed from its 
restriction to $\left({\cal T}({\cal Y}_j)\backslash {\cal Y}_j\right
)_{t=t^0}$.  But the identification (2.16)
shows that this holonomy is the same as the holonomy 
of the connection coming from theorem (1.26) on the
the space $D_j(a^0)\backslash \{a_j^0\}\times {\cal R}({\cal C})$.  Since $
\pi_1({\cal R}({\cal C}))=0$ we may
compute this holonomy by restricting to $\lambda =\lambda^0$ where
the connection agrees with $\bar{\nabla}^0$ by construction.

We also need to construct a model connection in the
tubular neighborhoods ${\cal T}({\cal Y}_j)\backslash {\cal Y}_j$ for $
j=m+1,\ldots ,n$ and
${\cal T}({\cal Y}_{\infty})\backslash {\cal Y}_{\infty}.$  In the first instance, we are looking for 
a connection on ${\cal T}({\cal Y}_j)$ with a simple type 0 singularity 
along ${\cal Y}_j$, 
which is equivalent to $\bar{\nabla}^0$ in a neighborhood of $a_j^
0$.  
Since $\bar{\nabla}$$^0$ is a simple type 0 connection, Proposition 1.25b
shows that in a neighborhood of $x=a_j^0$ there exists
a diagonal matrix $\Lambda_j$ so that in the appropriate local
trivialization about $x=a_j^0$, $\bar{\nabla}$$^0$ is represented by 
$$d_x+\Omega_j$$
on the trivial bundle 
$$D_j(a^0)\times {\bf C}^p\rightarrow D_j(a^0),\eqno (2.18)$$
where 
$$\Omega_j=-{{\Lambda_jd_x(x-a_j^0)}\over {x-a_j^0}}.$$
Define a 
connection $\nabla_j$  on the trivial bundle over ${\cal T}({\cal Y}_
j)\backslash {\cal Y}_j$ by
$$\nabla_j=d-{{\Lambda_jd(x-a_j(t))}\over {x-a_j(t)}},$$
where $d=d_x+d_t+d_{\lambda}$ is the exterior derivative on
${\cal T}({\cal Y}_j)$.  It is easy to check that $\nabla_j$ is a simple integrable
connection with type 0 singularity along ${\cal Y}_j$ which has a
restriction to $D_j(a^0)\times \{(t^0,\lambda^0)\}$ that is equivalent to $
\bar{\nabla}^0$ by
construction.
The same argument given above for $j=1,\ldots ,m$
applies here and one sees that the connection $\nabla_j$ defined
on the trivial bundle,
$${\cal T}({\cal Y}_j)\times {\bf C}^p\rightarrow {\cal T}({\cal Y}_
j),$$
has holonomy 
determined by its restriction to $t=t^0$ and $\lambda =\lambda^0$,
where it is essentially $\bar{\nabla}^0$.  To extend the connection
to ${\cal T}({\cal Y}_{\infty})\backslash {\cal Y}_{\infty}$ so that it is regular along $
{\cal Y}_{\infty}$, or has 
a logarithmic pole along ${\cal Y}_{\infty}$, or a higher rank simple
singularity one proceeds as above with some simplification
arising from the fact that $\infty$ adds no component to
the space of pole deformations.  One should use the
local parameter $w={1\over x}$ with
$$pr_{\infty}(w,t,\lambda )=(w,\lambda ).$$
We leave the details to the reader.

What we have now is a bundle $E_{\rho}$ over ${\bf P}^1\times {\cal D}
\backslash {\cal Y}$,
bundles $E_j$ over ${\cal T}({\cal Y}_j),$ and $E_{\infty}$ over $
{\cal T}({\cal Y}_{\infty})$ with
bundle isomorphisms, 
$$b_j:E_{\rho}|_{{\cal T}({\cal Y}_j)\backslash {\cal Y}_j}\rightarrow 
E_j|_{{\cal T}({\cal Y}_j)\backslash {\cal Y}_j},\eqno (2.19)$$
and
$$b_{\infty}:E_{\rho}|_{{\cal T}({\cal Y}_{\infty})\backslash {\cal Y}_{
\infty}}\rightarrow E_{\infty}|_{{\cal T}({\cal Y}_{\infty})\backslash 
{\cal Y}_{\infty}},\eqno (2.20)$$
which take the connection $\nabla_{\rho}$ into $\nabla_j$ and $\nabla_{
\infty}$.
We define a vector bundle $E$ over ${\bf P}^1\times {\cal D}$ by
forming the union $E_{\rho}\cup E_1\cup\cdots\cup E_n\cup E_{\infty}$ modulo
the equivalence relations determined by (2.19) and
(2.20).  By construction the bundle $E$ and the connection
$\nabla$ obtained by gluing together the connections $\nabla_{\rho}$, $
\nabla_j$, and
$\nabla_{\infty}$ have the properties asserted in the formulation
of Theorem 2.9.  This finishes the proof. QED
\vskip.2in \par\noindent
\S 3. {\bf Tau functions}
\vskip.2in \par\noindent
In this section we will first follow Helmink [11] to define
a tau function associated with the deformation 
construction of Theorem 2.0 above.  Let $(E,\nabla )$ be
the integrable deformation of $(E^0,\bar{\nabla}^0)$ constructed
in Theorem 2.9.  Recall that ${\cal D}={\cal R}(Z^n)\times {\cal R}
({\cal C})$ and define
$$\Theta :=\{(t,\lambda )\in {\cal D},\,\,E|_{{\bf P}^1\times \{(
t,\lambda )\}}\hbox{\rm is non-trivial}\}$$
\vskip.2in \par\noindent
{\bf Theorem 3.0}. {\sl There exists a non vanishing holomorphic map 
$\tau :{\cal D}\rightarrow {\bf C}$ so that $\Theta$ is equal to the zero set of $
\tau$. 
}     
\vskip.2in \par\noindent
This is basically a result of Malgrange [18] and
Helmink [11] which we will sketch a proof of following 
the arguments for Proposition 3.2 in [11].
\vskip.1in\par\noindent
Sketch of Proof (more details can be found in [11]).  Choose
$0<\rho_1<1<\rho_2$ and set
$$D_1:=\{x|x\in {\bf P}^1,|x|>\rho_1\},\,\,D_2:=\{x|x\in {\bf P}^
1,|x|<\rho_2\}.$$
Then $D_k\times {\cal D}$ for $k=1,2$ is a contractible Stein space, and  
hence $E|_{D_k\times {\cal D}}$ is holomorphically trivial for $k
=1,2$.
Let ${\bf f}:=(f_1,f_2,\ldots ,f_p)$ be a row vector of sections for
the restriction of $E$ to $D_1\times {\cal D}$ which trivializes this
restriction.  Let ${\bf g}:=(g_1,g_2,\ldots ,g_p)$ be a row vector of sections
for the restriction of $E$ to $D_2\times {\cal D}$ which trivializes this
restriction. Then there exists a holomorphic map 
$S:D_1\cap D_2\times {\cal D}\rightarrow GL(p,{\bf C})$ so that 
$${\bf g}={\bf f}S$$
on $D_1\cap D_2\times {\cal D}$.  Write $S(x,t,\lambda )=S_{t,\lambda}
(x)$.  Then the 
restriction of $E$ to ${\bf P}^1\times \{(t,\lambda )\}$ will be trivial if and only
if there are holomorphic maps $S_{t,\lambda}^{-}:D_1\rightarrow G
L(p,{\bf C})$ and
$S_{t,\lambda}^{+}:D_2\rightarrow GL(p,{\bf C})$ so that for all $
x\in D_1\cap D_2$,
$$S_{t,\lambda}(x)=S_{t,\lambda}^{-}(x)S_{t,\lambda}^{+}(x)^{-1}.\eqno 
(3.1)$$
A $p$ vector of sections that trivializes $E$ is then given
by the appropriate extension of 
$${\bf g}S_{t,\lambda}^{+}={\bf f}S_{t,\lambda}^{-},$$
from $D_1\cap D_2$.  Let $S^1$ denote the unit disk, write ${\cal H}
:=L^2(S^1,{\bf C}^p)$ 
and let ${\cal H}_{+}$ be the closed subspace of ${\cal H}$ consisting of 
those functions with are boundary values functions 
holomorphic inside the unit disk.  Let ${\cal H}_{-}={\cal H}_{+}^{
\perp}.$  Suppose
that $S^1\ni x\rightarrow S(x)$ is a smooth $p\times p$ matrix valued function on
the circle and let ${\cal H}\ni f\rightarrow Sf$ denote the associated 
multiplication operator on ${\cal H}$.  Let 
$$S=\left[\matrix{a(S)&b(S)\cr
c(S)&d(S)\cr}
\right],$$
denote the decomposition of $S$ relative to the direct sum
decomposition ${\cal H}={\cal H}_{+}\oplus {\cal H}_{-}$.  It is well known [1] that the 
factorization (3.1) exists if and only if 
$a(S_{t,\lambda}):{\cal H}_{+}\rightarrow {\cal H}_{+}$ is invertible.  Since $
(t,\lambda )\rightarrow a(S_{t,\lambda})$ is
continuous in the uniform norm topology for $a(S_{t,\lambda})$ 
and $a(S)$ is known to be Fredholm if $S$ is smooth,
it follows that the index of $a(S_{t,\lambda})$ is independent of
$(t,\lambda ).$ Since $a(S_{t^0,\lambda^0})$ is invertible by construction it
follows that the index of $a(S_{t,\lambda})$ is 0 for all $(t,\lambda 
)\in {\cal D}$.
Fix $(t,\lambda )$ for the moment.  Then since $a(S_{t,\lambda})$ has index
0, there exists a finite rank operator $k:{\cal H}_{+}\rightarrow 
{\cal H}_{+}$ so that
$k+a(S_{t,\lambda})$ is invertible. In fact there exists a 
neighborhood $V_{t,\lambda}$ of $(t,\lambda )$ so that for all $(
s,\mu )\in V_{t,\lambda}$ the
operator $q_{s,\mu}:=k+a(S_{s,\mu})$ is invertible.  Note that
$q_{s,\mu}$ is a parametrix for $a(S_{s,\mu})$ for $(s,\mu )\in V_{
t,\lambda}$ in
that
$$a(S_{s,\mu})q_{s,\mu}^{-1}=I+F_{s,\mu}$$
where $F_{s,\mu}$ is a finite rank operator that depends 
holomorphically on $(s,\mu ).$  Now we show that it is
possible to make a coherent choice of parametrices
$q_{s,\mu}$ following the argument in Helmink [11]. Let $\{V_i\}$ be
a locally finite covering of ${\cal D}$ with $q_i(s,\mu )$ a holomorphic 
parametrix
for $a(S_{s,\mu})$ for all $(s,\mu )\in V_i.$  For each $i,j$ such that
$V_i\cap V_j\ne\phi$ define $q_iq_j^{-1}=\phi_{ij}$.  Note that each $
\phi_{ij}$ is
a finite rank perturbation of the identity.  The 
maps $V_i\cap V_j\ni (s,\mu )\rightarrow\det\phi_{ij}(s,\mu )$ are the transition
functions for a holomorphic line bundle on ${\cal D}$.  Since
${\cal D}$ is a Stein space we have $H^1({\cal D},{\cal O}^{*})=0$, and it follows
that this line bundle must be trivial.  Hence there exist
holomorphic maps $\tau_i:V_i\rightarrow {\bf C}^{*}$ so that $\tau_
i^{-1}\tau_j=\det\phi_{ij}$.
Now define $t_i:{\cal H}_{+}\rightarrow {\cal H}_{+}$ by $t_i1=\tau_
i$ and $t_i(e^{in\theta})=e^{in\theta}$ for
$n=1,2,\ldots$.  Now define ${\bf q}_i=t_iq_i$.  Then ${\bf q}_i(
s,\mu )$ remains
a holomorphic parametrix for $a(S_{s,\mu})$ for $(s,\mu )\in V_i$ and
since
$$\det {\bf q}_i{\bf q}_j^{-1}=\det t_it_j^{-1}\det q_iq_j^{-1}=\tau_
i\tau_j^{-1}\det\phi_{ij}=1,$$
it follows that 
$$\tau (s,\mu ):=\det(a(S_{s,\mu}){\bf q}_i(s,\mu )^{-1})$$
is a well defined holomorphic function on all of ${\cal D}$ whose
0 set is equal to $\Theta$. QED

Next will make a connection with the tau function 
defined by Jimbo, Miwa and Ueno [12].  We follow 
Malgrange by computing the
regularized logarithmic derivative of the determinant of
a Toeplitz operator, the invertibility of which 
determines if $E|_{{\bf P}^1\times \{(t,\lambda )\}}$ is trivial or not.  The bundle 
$E$ is first realized in terms of a system of transition 
functions which relate local models for the connection 
$\nabla$ in a neighborhood of the singularities to a model for
the connection $\nabla$ in a complement of a neighborhood of the
singularities.  We will spend some effort to choose the
local models carefully (so that (3.7) below is satisfied) 
even though this is not  important for the calculation
of the regularized logarithmic derivative of the Toeplitz
operator.  We do it because it will simplify
a curvature calculation later on.

First fix a
point $(t^0,\lambda^0)\in {\cal D}$.  We will exam the restriction of $
E$ to
${\bf P}^1\times W$ where $W$ is a suitably small neighborhood of 
$(t^0,\lambda^0).$ We will cover ${\bf P}^1\times W$ by neighborhoods 
$B_j\times W$ which contain the singular sets $x=a_j(t)$ 
(including $x=\infty$ for $j=\infty$) and a complementary set
$B_{ex}\times W$.  We will choose trivializations for $E$ over 
each of these sets and understand the bundle $E$ in terms
of the transition maps between these trivializations.

Now we turn to the identification of a suitable local model
for the connection $\nabla$ in a neighborhood of each singularity.

For each
$j=1,\ldots ,m$ choose a connected,
simply connected product neighborhood $W_j=U_j\times V_j$ 
of $(t^0_j,\lambda_j^0)$, with compact closure.  For $j=m+1,\ldots 
,n$ 
($j=\infty )$ choose a connected, simply connected neighborhood of 
$t^0_j$ $(\lambda^0_{\infty})$ with compact closure.  Let   
$$W=\prod_{j=1}^nW_j\times W_{\infty}.$$

Let $D_j(a^0)$ denote the disk of radius $\delta (a^0)/3$ defined in
section 2. Choose a trivialization, ${\bf g}^j,$
for the restriction $E|_{D_j(a^0)\times W}$ and suppose that relative to 
this trivialization the
connection $\nabla$ is given by
$$d+\Omega_j.$$
Let $\Omega_j^0$ denote the restriction of $\Omega_j$ to $(t_j,\lambda_
j)=(t_j^0,\lambda_j^0)$. 
As in the developments preceeding Theorem 1.35 it is 
possible to adjust the trivialization ${\bf g}^j$ by a gauge 
transformation in the $x$ variables alone so that $d_x+\Omega_j^0$ extends
to a connection on the trivial bundle ${\bf P}^1\times {\bf C}^p\rightarrow 
{\bf P}^1$ with a
simple type $r$ singularity at $a_j^0$ and a regular singularity
at $\infty$.  In what follows we suppose that this has been 
done.
For $j=1,\ldots ,m$ choose $V_j$ small enough so that the Birkhoff deformation
of $d_x+\Omega_j^0$ constructed in Proposition 1.35 exists on
$D_j(a^0)\times V_j$.  Thus there exists an integrable connection
$$d_x+d_{\lambda_j}+\Omega_j^{loc},\eqno (3.2)$$
which is a Birkhoff deformation of $d_x+\Omega^0_j$ in the sense of 
Proposition 1.35 (i) and (ii) and such that on 
$D_j(a^0)\backslash \{a_j^0\}\times V_j$ one has the gauge equivalence
$$d_x+d_{\lambda_j}+\Omega_j^{loc}=\varphi_j\cdot\left[d_x+d_{\lambda_
j}+\Omega^0_j\right]\eqno (3.3)$$
where $x\rightarrow\varphi_j(x,\lambda_j)$ is holomorphic in the punctured  
disk $D_j(a^0)\backslash \{a_j^0\}$ and asymptotic to the identity as
$x\rightarrow\infty$.  Now
choose the neighborhood $U_j$ of $\{t_j^0\}$ small enough so that
the map, $p_j$ defined by
$$D_j(a^0)\times U_j\times V_j\ni (x,t,\lambda )\rightarrow p_j(x
,t,\lambda ):=(x-a_j(t)+a_j^0,\lambda )$$
maps into $D_j(a^0)\times V_j.$  Let $d_j=d_x+d_{t_j}+d_{\lambda}$, where
$d_{\lambda}$ is the exterior derivative on $\prod_{k=1}^n{\cal R}
({\cal C}_k)\times {\cal R}({\cal C}_{\infty})$ and define 
the connection
$$d_j+p_j^{*}\Omega_j^{loc},\eqno (3.4)$$
on the trivial ${\bf C}^p$ bundle over 
$$X_j:=D_j(a^0)\times W.$$
It is a consequence of Proposition 1.26c above  
that the 
connection (3.4) on the trivial 
bundle over $X_j$ is equivalent to $\nabla$ on $E|_{X_j}$ (they have the
same formal reduction and Stokes' multipliers).
Possibly
shrinking the open neighborhood $U_j$ we can insure that
there is an annular region  
$$A_j=\{x|\rho_j<|x-a_j^0|<\rho'_j\}\subset D_j(a^0)$$
with the property that $x-a_j(t)\ne 0$ for 
$(x,t)\in A_j\times U_j$.  The integrable connection (3.4) and
the integrable connection
$$d_j+\Omega_j^{loc}$$
have the same holonomy on $A_j\times W_j$.  They are
thus related by a holomorphic gauge transformation, $\psi_j,$
defined on $A_j\times W_j$ so that
$$d_j+p_j^{*}\Omega_j^{loc}=\psi_j\cdot\left[d_j+\Omega_j^{loc}\right
],\eqno (3.5)$$
which can be normalized so that $\psi_j(x,t^0_j,\lambda_j^0)=I$ for
$x\in A_j.$ By choosing $W_j$ small enough we may insure
that $\psi_j$ is a sufficiently small perturbation of the 
identity so that it has a ``canonical factorization''
$\psi_j=\psi_j^{+}\psi_j^{-}$, where $\psi_j^{+}(x,t_j,\lambda_j)$ is holomorphic for 
$|x-a_j^0|<\rho'_j$ and $\psi_j^{-}(x,t_j,\lambda_j)$ is 
holomorphic for $|x-a_j^0|>\rho_j$, 
with $\psi^{-}_j(\infty ,t_j,\lambda_j)=I$.  Now define the 
gauge transform of $d_j+p^{*}_j\Omega_j^{loc}$
by $\left(\psi_j^{+}\right)^{-1}$,
$$\nabla_j:=\left(\psi^{+}\right)^{-1}\cdot\left[d_j+p^{*}_j\Omega_
j^{loc}\right]=\psi_j^{-}\cdot\left[d_j+\Omega_j^{loc}\right].\eqno 
(3.6)$$
Combining (3.6) with the extension of (3.3) to

$$d_j+\Omega_j^{loc}=\varphi_j\cdot\left[d_j+\Omega^0_j\right],$$
which follows from (3.3) since $\varphi_j$ does not depend on 
$t_j$, or $\lambda_k$ for $k\ne j$ we also have,
$$\nabla_j=\phi_j\cdot\left[d_j+\Omega_j^0\right],\eqno (3.7)$$
where $\phi_j(x,t_j,\lambda_j)$ is holomorphic for $|x-a_j^0|>\rho_
j.$  The
connection $\nabla_j$ is equivalent to the restriction of $\nabla$
to $E|_{X_j}$ and so is a good local model for $\nabla$.  Thus we can
choose a trivialization ${\bf f}^j$ for $E|_{X_j}$ so that in this 
trivialization $\nabla$ is represented by $\nabla_j$.

 Now we wish to do something similar for
$j=m+1,\ldots ,n$.  
$W_j$ should be small enough so the the map 
$$D_j(a^0)\times W_j\ni (x,t)\rightarrow p_j(x,t):=x-a_j(t)+a_j^0
,$$
maps into $D_j(a^0)$.  Choose a trivialization for 
$E|_{D_j(a^0)\times W}$  and write $\Omega_j$ for the connection one
form for $\nabla$ in this trivialization.  Thus $\nabla$ is
represented by
$$d+\Omega_j,$$
in this trivialization.

Let $\Omega_j^0$ denote the restriction of $\Omega_j$ to
$(t,\lambda )=(t^0,\lambda^0)$.  Again using the same argument to
be found in the preliminaries to Proposition 1.35 we
may suppose that the trivialization of $E|_{D_j(a^0)\times W}$ has
been chosen so that $d_x+\Omega_j^0$ extends to a meromorphic
connection on the trivial bundle ${\bf P}^1\times {\bf C}^p\rightarrow 
{\bf P}^1$ with a simple
type 0 singularity at $x=a$ and a regular singularity at
$x=\infty$.  

As above consider the connection
$$d_j+p_j^{*}\Omega_j^0.$$
One can choose $W_j$ small enough so that there
exists an annular region $A_j=\{x|\rho_j<|x-a_j^0|<\rho'_j\}$ 
contained in $D_j(a^0)$ with the property that $x-a_j(t)\ne 0$
for $(x,t_j)\in A_j\times W_j.$  The connection $d_j+p_j^{*}\Omega_
j^0$ and
$d_j+\Omega_j^0$ then have the same holonomy over $A_j\times W$
and so there exists a holomorphic gauge transformation
$\psi_j$ defined on $A_j\times W$ so that
$$d_j+p^{*}_j\Omega_j^0=\psi_j\cdot\left[d_j+\Omega_j^0\right].$$
One can normalize $\psi_j$ so that $\psi_j(x,t^0,\lambda^0)=I$ and
by choosing $W$ sufficiently small we can guarentee that
$\psi_j$ has a canonical factorization $\psi_j^{+}\psi_j^{-}$ as above.  We
define
$$\nabla_j:=\left(\psi_j^{+}\right)^{-1}\cdot\left[d_j+p_j^{*}\Omega_
j^0\right]=\psi_j^{-}\cdot\left[d_j+\Omega_j^0\right],\eqno (3.8)$$
where $\psi_j^{-}(x,t,\lambda )$ is holomorphic for $|x-a_j^0|>\rho_
j$ and
asymptotic to the identity $I$ as $x\rightarrow\infty$.  As above we can
find a trivialization ${\bf f}^j$ for the restriction of $E$ to 
$D_j(a^0)\times W$ so that $\nabla$ is represented by $\nabla_j$.
$$ $$

The 
local connection at infinity, $\nabla_{\infty},$ might be regular, or
have a 
simple type $r_{\infty}$ singularity.  It does not 
matter much for the calculation we are going to do, but
for definiteness we suppose $r_{\infty}\ge 1.$  Then $W_{\infty}$ is a 
neighborhood of $\lambda_{\infty}^0$ (no pole deformation parameters).
Proceeding in close analogy with the first case discussed
above we choose  a trivialization ${\bf g}^{\infty}$ for
$E$ restricted to $D_{\infty}(a^0)\times W$ so that the connection 
$\nabla$ is represented by
$$d_x+d_{\lambda}+\Omega_{\infty}.\eqno $$
We further suppose that the trivialization ${\bf g}^{\infty}$ has been 
chosen so that the restriction of this connection to
$\lambda =\lambda^0$, given by $d_x+\Omega_{\infty}^0$, extends to a connection on the
trivial bundle ${\bf P}^1\times {\bf C}^p\rightarrow {\bf P}^1$ with a simple type $
r_{\infty}$  
singularity at $\infty$ and a regular singular point at 0.
 
Now let 
$$d_x+d_{\lambda_{\infty}}+\Omega_{\infty}^{loc}$$
denote the Birkhoff deformation of $d_x+\Omega_{\infty}^0$ constructed 
in Proposition 1.35 (with the slight alterations needed to 
locate the singularity at $\infty$).  Let $p_{\infty}$ denote the 
projection $(x,\lambda )\rightarrow (x,\lambda_{\infty})$ and define
$$\nabla_{\infty}:=d_x+d_{\lambda}+p_{\infty}^{*}\Omega_{\infty}^{
loc}.\eqno (3.9)$$
Choose $\rho_{\infty}$ and $\rho'_{\infty}$ so
that $\nabla_{\infty}$ and $d_x+d_{\lambda}+\Omega_{\infty}^0$ are holomorphically equivalent 
on the annulus
 
$$A_{\infty}=\{x|\rho_{\infty}<x<\rho'_{\infty}\}\subset D_{\infty}
(a^0),$$
by a gauge transformation $\phi_{\infty}$,
$$\nabla_{\infty}=\phi_{\infty}\cdot\left[d_x+d_{\lambda}+\Omega^
0_{\infty}\right],\eqno (3.10)$$
which can be chosen so that $\phi_{\infty}(x,\lambda )$ is holomorphic for $
|x|<\rho'_{\infty}.$

In contrast Theorem 2.9 we do {\it not\/} assume that
the restriction
$$E|_{{\bf P}^1\times \{t^0,\lambda^0\}}$$
is holomorphically trivial.  Let $B_j(\rho )=\{x:|x-a_j^0|<\rho \}$ 
denote the open ball of radius $\rho$ about $a_j^0$.  Let 
$B_{\infty}(\rho )=\{x:|x|>\rho \},$ 
and define
 
$$\eqalign{B_j:=&B_j(\rho'_j),\cr
B_{\infty}:=&B_{\infty}(\rho_{\infty}),\cr
B:=&B_1(\rho_1)\cup B_2(\rho_2)\cup\cdots\cup B_n(\rho_n)\cup B_{
\infty}(\rho'_{\infty}),\cr
B_{ex}:=&{\bf P}^1\backslash\bar {B},\cr}
\eqno (3.11)$$
where $\bar {X}$ denotes the closure of $X$. Then
$$\{B_1,B_2,\ldots ,B_n,B_{\infty},B_{ex}\}\eqno (3.12)$$
is an open covering of ${\bf P}^1$. 

 We now wish to show that 
there is a holomorphic trivialization of the bundle $E$ 
over $B_{ex}\times W$.  Since we've seen that $\Theta$ is the zero set
of a nonvanishing holomorphic function, $\tau ,$ it follows that 
there is some $(t^1,\lambda^1)\in W$ so that $E|_{{\bf P}^1\times 
\{(t^1,\lambda^1)\}}$ is 
holomorphically trivial.
Since $B_{ex}\times W$ does not intersect any of the singular sets
(3.2) or $\{\infty \}\times W$ it follows that the integrable connection
$\nabla$ on $E$ is smooth over $B_{ex}\times W$, and since $W$ is connected 
and simply
connected one may integrate $\nabla$ over $W$ to extend the
trivialization over $B_{ex}\times \{(t^1,\lambda^1)\}$ to a holomorphic 
trivialization of $E$ over $B_{ex}\times W.$ We wish to pick a 
trivialization so that the connection form for $\nabla$ is 
particularly simple.  Suppose that relative to some 
choice of a trivialization for the restriction of $E$ to
${\bf P}^1\times \{(t^1,\lambda^1)\}$ the connection $\nabla$ is,
$$d_x+\Omega_{ex}(x),$$
where $\Omega_{ex}(x)$ depends only on $x$.
Write $d=d_x+d_t+d_{\lambda}$; then since $W$ is connected and 
simply connected it is
easy to see that the connection,
$$d+\Omega_{ex}(x),\eqno (3.13)$$
defined on the trivial bundle over $B_{ex}\times W$ with fiber
${\bf C}^p$ has the same holonomy representation as $\nabla$ on
the restriction of $E$ to $B_{ex}\times W$. 
The holomorphic
equivalence of these two connections implies that we
can choose a trivialization ${\bf f}^{ex}$ for
$E|_{B_{ex}\times W}$ so that relative to this trivialization the 
connection $\nabla$ is given by (3.13).  Observe that this 
connection form has no $dt$ or $d\lambda$ components.
$$ $$

For each $j=1,\ldots ,n$ there 
exists a holomorphic map $S^j:A_j\times W\rightarrow GL(p,{\bf C}
)$ so that
$${\bf f}^{ex}={\bf f}^jS^j,\eqno (3.14)$$
and a holomorphic map $S^{\infty}:A_{\infty}\times W\rightarrow G
L(p,{\bf C})$ so 
that 
$${\bf f}^{ex}={\bf f}^{\infty}S^{\infty},\eqno (3.15)$$
where we think of each trivialization ${\bf f}=(f_1,\ldots ,f_p)$ as
a row vector of sections.

Define 
$$B_{in}=\cup_jB_j\cup B_{\infty},$$
 Let $S$ denote the 
holomorphic map from $B_{in}\cap B_{ex}\times W$ into $GL(p,{\bf C}
)$ which 
restricts to $S^j$ on $A_j\times W$ and $S^{\infty}$ on $A_{\infty}
\times W$.  Then for $(t,\lambda )\in W,$
$E|_{{\bf P}^1\times \{(t,\lambda )\}}$ will be holomorphically trivial if and only if 
there exists a factorization,
$$S(x,t,\lambda )=\Phi_{in}(x,t,\lambda )^{-1}\Phi_{ex}(x,t,\lambda 
),\eqno (3.16)$$
where $x\rightarrow\Phi_{in}(x,t,\lambda )$ is holomorphic and invertible in $
B_{in}$
and $x\rightarrow\Phi_{ex}(x,t,\lambda )$ is holomorphic and invertible in $
B_{ex}$.
Let $C_j$ $(C_{\infty})$ denote a counterclockwise oriented circle
contained in $A_j$ ($A_{\infty}$), and define the
oriented curve  
$$C=C_{\infty}-C_1-C_2-\cdots -C_n.$$
Choose a point $z_0\in B_{ex}$ with $z_0\notin\overline {B_{in}}$ and define a 
projection $P_{ex}$ on $L^2(C)$ by,
$$P_{ex}f(z)=\int_Cf(x)\left\{{1\over {x-z}}-{1\over {x-z_0}}\right
\}{{dx}\over {2\pi i}}.$$
This projects onto the subspace of $L^2(C)$ which has
a holomorphic extension into the connected part of
$B_{ex}$ bounded by the curve $C$, with the further property
that this holomorphic extension vanishes at $z=z_0$.
We can make the factorization (3.6) unique (when it
exists) by normalizing 
$$\Phi_{ex}(z_0,t,\lambda )=\hbox{\rm identity}=I.$$
Rewriting (3.16) we have
$$\Phi_{in}S=\Phi_{ex},$$
and writing $P_{in}=I-P_{ex}$ we find
$$P_{in}(\Phi_{in}S)=I.\eqno (3.17)$$
Regarding this as an equation for the rows of $\Phi_{in}$, the
solution of this equation is equivalent to the existence 
of the factorization (3.16).  Suppose that (3.16) has a 
solution $\Phi_{in}$, $\Phi_{ex}$.  Then the Toeplitz operator $T_
S$ 
defined by
$$T_Sf=P_{in}(fS),$$
where $f$ is a row vector, has inverse,
$$T_S^{-1}g=P_{in}(g\Phi_{ex}^{-1})\Phi_{in}.$$
Following Malgrange we now calculate the regularized 
trace
$$\omega :=\hbox{\rm Tr}\left(T_S^{-1}T_{dS}-T_{dSS^{-1}}\right),\eqno 
(3.18)$$
where for brevity we write $d=d_{t,\lambda}$. 
\vskip.1in\par\noindent
Note: because the multiplication 
operator in our Toeplitz operator is acting on the 
right it is $T_{RS}-T_ST_R$ which is compact when $R$ and
$S$ are smooth.  
\vskip.1in\par\noindent
We will eventually show that (3.18) 
differs from
$d\log\tau$, defined above, by a regular term and this will 
allow us to make a connection with the formula for
the tau function
given by Jimbo, Miwa and Ueno in [12].  

We compute
$$\eqalign{T_S^{-1}T_{dS}f&=P_{in}\left(\left(fdS\right)\Phi_{ex}^{
-1}\right)\Phi_{in}\cr
&=P_{in}\left(fdS\Phi_{ex}^{-1}\right)\Phi_{in}\cr
&=P_{in}\left(f\left(\Phi_{in}^{-1}d\Phi_{ex}\Phi_{ex}^{-1}-\Phi_{
in}^{-1}d\Phi_{in}\Phi_{in}^{-1}\right)\right)\Phi_{in},\cr}
\eqno (3.19)$$
and
$$\eqalign{T_{dSS^{-1}}f&=P_{in}\left(fdSS^{-1}\right)\cr
&=P_{in}\left(f\left(\Phi_{in}^{-1}d\Phi_{ex}\Phi_{ex}^{-1}\Phi_{
in}-\Phi_{in}^{-1}d\Phi_{in}\right)\right).\cr}
\eqno (3.20)$$
Now let
$$R_qf=fq,$$
denote right multiplication by $q$, and define
$$Q=\Phi_{in}^{-1}d\Phi_{ex}\Phi_{ex}^{-1}-\Phi_{in}^{-1}d\Phi_{i
n}\Phi_{in}^{-1}.$$
Then from (3.9) and (3.10) one sees that
$$\eqalign{T_S^{-1}T_{dS}-T_{dSS^{-1}}=&R_{\Phi_{in}}P_{in}R_Q-P_{
in}R_{\Phi_{in}}R_Q\cr
=&\left[R_{\Phi_{in}},P_{in}\right]R_Q\cr
=&\left[P_{ex},R_{\Phi_{in}}\right]R_Q.\cr}
$$
Thus the trace of interest is
$$\hbox{\rm Tr}\left[P_{ex},R_{\Phi_{in}}\right]R_Q\eqno (3.21)$$
and one computes
$$\eqalign{\left[P_{ex},R_{\Phi_{in}}\right]R_Qf(z)&=\cr
\int_Cf(x)Q(x)&(\Phi_{in}(x)-\Phi_{in}(z))\left\{{1\over {x-z}}-{
1\over {x-z_0}}\right\}{{dx}\over {2\pi i}}.\cr}
$$
Writing $d_x\Phi =\Phi'dx$, the integral of the (finite dimensional) trace 
over the diagonal is then
$$\eqalign{\int_C\hbox{\rm Tr}\left(Q(x)\Phi_{in}'(x)\right){{dx}\over {
2\pi i}}&\cr
=\int_C\hbox{\rm Tr}&\left(d\Phi_{ex}\Phi_{ex}^{-1}\Phi_{in}'\Phi_{
in}^{-1}-d\Phi_{in}\Phi_{in}^{-1}\Phi_{in}'\Phi_{in}^{-1}\right){{
dx}\over {2\pi i}}\cr
=\int_C\hbox{\rm Tr}&\left(d\Phi_{ex}\Phi_{ex}^{-1}\Phi_{in}'\Phi_{
in}^{-1}\right){{dx}\over {2\pi i}},\cr}
\eqno (3.22)$$
where the second equality follows from Cauchy's 
theorem and the fact that $\Phi_{in}$ is holomorphic in $B_{in}$.
To make a connection between (3.22) and the JMU [12]
formula for the log derivative of the tau function we
replace $\Phi_{in}$ and $\Phi_{ex}$ in (3.22) with quantities more 
intimately related to the connection $\nabla$.  First choose
$j\in \{1,\ldots ,n,\infty \}$, and let $\nabla_j$ denote the representation for
$\nabla$ in the trivialization ${\bf f}^j$.  Let $\nabla_{ex}$ denote the 
representation of $\nabla$ in the trivialization ${\bf f}^{ex}$. 
Then
in the trivialization ${\bf f}^j\Phi_{in}^{-1}={\bf f}^{ex}\Phi_{
ex}^{-1}$ for the restriction
of $E$ to ${\bf P}^1\times\left(W\backslash\Theta\right)$ we find that the connection $
\nabla$ is
given by,
$$\nabla =\Phi_{ex}\left[\nabla_{ex}\right]=\Phi_j\cdot\left[\nabla_
j\right]\eqno (3.23)$$
on $A_j\times W\backslash\Theta$ (or $A_{\infty}\times W\backslash
\Theta$ for $j=\infty$).  

Now fix $j\in \{1,\ldots ,m,\infty \}$ and let
$\hat{\alpha}_j$ be the formal power series near $x=a_j(t)$ (or $
\infty$ if 
$j=\infty$) for which
$$\hat{\alpha}_j\cdot\left[\nabla_{\lambda_j}\right]=\nabla\eqno 
(3.24)$$
Let $\hat{\alpha}_j(loc)$ denote the formal power series near $x=
a_j(t)$
such that
$$\hat{\alpha}_j(loc)\cdot\left[\nabla_{\lambda_j}\right]=\nabla_
j\eqno (3.25)$$

Equating (3.24) with the first term of (3.23) and recalling
that $\Omega_{ex}$ does not have any $dt$ or $d\lambda$ terms, one finds
$$d\Phi_{ex}\Phi_{ex}^{-1}=d\hat{\alpha}_j\hat{\alpha}_j^{-1}-\hat{
\alpha}_jdH_j\hat{\alpha}_j^{-1},\eqno (3.26)$$
which should be understood in the following sense.  
Replacing $\hat{\alpha}_j$ by $\alpha_{\Sigma ,j}$ defined in a suitable sector $
\Sigma$ with
vertex at $x=a_j(t)$ one finds a sectorial version of (3.26).
This shows that the function $d\Phi_{ex}\Phi_{ex}^{-1}$ which is 
analytic in an annular region about $x=a_j(t)$ extends holomorphically
into the sector $\Sigma$. Since this is true for a collection of 
sectors which cover a punctured neighborhood of 
$x=a_j(t)$ it follows that $d\Phi_{ex}\Phi_{ex}^{-1}$ is holomorphic in a 
punctured neighborhood of $x=a_j(t)$.  Equation (3.26) may 
then be understood as an equality of formal Laurent 
series (which in fact converge since the left hand side
has a convergent Laurent series).

Applying $\hat{\Phi}_{in}$ (the formal power series associated to $
\Phi_{in}$) 
to both sides of (3.25) and comparing the
result with (3.23) and (3.24) one finds 
$$\hat{\Phi}_{in}\hat{\alpha}_j(loc)=\hat{\alpha}_jc_j$$
where $c_j$ is a diagonal constant matrix (the only 
gauge automorphisms
of $\nabla_{\lambda_j}$ are diagonal constants).  Thus 
$$\hat{\Phi}_{in}=\hat{\alpha}_jc_j\hat{\alpha}_j(loc)^{-1},\eqno 
(3.27)$$
for $j=1,2,\ldots ,m,\infty$.  This is to be understood in the 
sense of formal power series at $x=a_j(t)$.

For $j=m+1,\ldots ,n$ let $\alpha_j$ denote the
local holomorphic gauge transformation 
constructed in Proposition 1.25b such that
$$\alpha_j\cdot\left[d_x+d-{{\Lambda_j}\over {x-a_j(t)}}d(x-a_j(t
))\right]=\nabla ,\eqno (3.28)$$
where $\Lambda_j$ is a constant diagonal matrix.  Let $\alpha_j(l
oc)$ denote the
local holomorphic gauge transformation so that
$$\alpha_j(loc)\cdot\left[d_x+d-{{\Lambda_j}\over {x-a_j(t)}}d(x-
a_j(t))\right]=\nabla_j.\eqno (3.29)$$
$$ $$
Comparing this with the first term in (3.23) and making
use of the fact that $\Omega_{ex}$ has no $dt$ or $d\lambda$ components
one finds
$$d\Phi_{ex}\Phi_{ex}^{-1}=d\alpha_j\alpha_j^{-1}+\alpha_j{{\Lambda_
jda_j(t)}\over {x-a_j(t)}}\alpha_j^{-1}.\eqno (3.30)$$
If we write $dH_j={{\Lambda_jd(x-a_j(t))}\over {(x-a_j(t))}}$ for $
j=m+1,\ldots ,n$, 
(remember $d=d_t+d_{\lambda}$) then
(3.30) can be written 
$$d\Phi_{ex}\Phi_{ex}^{-1}=d\alpha_j\alpha_j^{-1}-\alpha_jdH_j\alpha_
j^{-1},\eqno (3.31)$$
which is  analogous to (3.26).

Applying $\Phi_{in}$ to both sides of (3.29) and comparing with 
(3.28) we find
$$\Phi_{in}=\alpha_jc_j\alpha_j(loc)^{-1}\hbox{\rm \ for }j=m+1,\ldots 
,n\eqno (3.32)$$
which is  analogous to (3.27). 

The formal power series expansion for (3.26) and the 
powers series expansion for (3.30) shows that
the integral,  
 
$$\int_{C_j}\hbox{\rm Tr}\left(d\Phi_{ex}\Phi_{ex}^{-1}\Phi'_{in}
\Phi_{in}^{-1}\right){{dx}\over {2\pi i}},$$
can be ``done'' by residues to get,
$$\hbox{\rm $\pm$Res}_{x=a_j(t)}\hbox{\rm Tr}\left(d\Phi_{ex}\Phi_{
ex}^{-1}\Phi'_{in}\Phi_{in}^{-1}\right),\eqno (3.33)$$
with the $+$ choice for $j=1,\ldots ,n$ and the - choice for
$j=\infty$. We now substitute (3.26), (3.27), (3.31) and (3.32) into 
(3.33).  The first 
term in (3.26) and (3.30) does not make a contribution to the formal
residue in (3.28) since $d\hat{\alpha}_j\hat{\alpha}_j^{-1}$ and $
\Phi'_{in}\Phi_{in}^{-1}$ both have 
formal power series expansions at $x=a_j(t)$. After some
simplification (using $c_j^{-1}dH_jc_j=dH_j$) one finds
$$\eqalign{\int_{C_j}\hbox{\rm Tr}\left(d\Phi_{ex}\Phi_{ex}^{-1}\Phi'_{
in}\Phi_{in}^{-1}\right)&{{dx}\over {2\pi i}}\cr
=\pm\hbox{\rm Res}_j\hbox{\rm Tr}&\left(\hat\alpha_j^{-1}\hat\alpha_
j'dH_j-\hat\alpha_j(loc)^{-1}\hat\alpha_j'(loc)dH_j\right),\cr}
\eqno (3.34)$$
where $\hbox{\rm Res}_j$ is the residue at $x=a_j(t)$ or $\infty$ if $
j=\infty$,
and the sign $\pm$ is $+$ for $j=1,\ldots ,n$ and - for $j=\infty$.  
Combining (3.22) with (3.34) one finds that
$$\eqalign{\hbox{\rm Tr}\left(T_S^{-1}T_{dS}-T_{dSS^{-1}}\right)&\cr
=-\sum_j\hbox{\rm Res}_j\hbox{\rm Tr}&\left(\hat\alpha_j^{-1}\hat
\alpha_j'dH_j-\hat\alpha_j(loc)^{-1}\hat\alpha_j'(loc)dH_j\right)\cr}
\eqno (3.35)$$
where the sum is over $j\in \{1,\ldots ,n,\infty \}$.  

Next we make use of the special choice we made for 
the local models $\nabla_j$ that is reflected in (3.7), (3.8) and (3.10).
Following the arguments for the proof of part (iv) of
Proposition 1.35 we find that
$$d_{t,\lambda}\hbox{\rm Res}_j\hbox{\rm Tr}\left(\hat\alpha_j(lo
c)^{-1}\hat\alpha_j'(loc)dH_j\right)=0,\eqno (3.36)$$
for $j=1,\ldots ,n,\infty$ follows from (3.7), (3.8), and (3.10) in
the same way that (1.42) follows from (1.41).

We will now make use of (3.36) to show that the one form in 
(3.35) is closed off the singular set $\Theta$.  One easily
computes 
$$\eqalign{d\hbox{\rm Tr}&\left(T_S^{-1}T_{dS}-T_{dSS^{-1}}\right
)\cr
=-{1\over 2}\sum_{j>k}\hbox{\rm Tr}&\left(\left[T_S^{-1}T_{\partial_
kS},T_S^{-1}T_{\partial_jS}\right]+T_{\left[\partial_jSS^{-1},\partial_
kSS^{-1}\right]}\right)ds_j\wedge ds_k\cr}
,$$
where $s:=(t,\lambda )$ and $\partial_k={{\partial}\over {\partial 
s_k}}$.
This last expression can be computed as in Malgrange [17] 
and one finds
$${1\over 2}\sum_{j>k}\int_C{{dx}\over {2\pi i}}\hbox{\rm Tr}\left
(\partial_jSS^{-1}\left(\partial_kSS^{-1}\right)'\right)ds_j\wedge 
ds_k.\eqno (3.37)$$
If $(t^0,\lambda^0)\notin\Theta$ then we may take $(t^1,\lambda^1
)=(t^0,\lambda^0)$ in the
construction above.  Note that equation (3.35) shows
that $\hbox{\rm $\hbox{\rm Tr}\left(T_S^{-1}T_{dS}-T_{dSS^{-1}}\right
)$ }$is actually independent of the
choice of $(t^1,\lambda^1)$, though it does depend on $(t^0,\lambda^
0)$ through
$\hat{\alpha}_j(loc)$.  With this choice for $(t^1,\lambda^1)$ equations (3.7), (3.8)
and (3.10) take on added significance.  In this case one 
may choose a global trivialization for $E|_{{\bf P}^1\times \{(t^
0,\lambda^0)\}}$.  If
$\Omega_j^0$ denotes the one form for $\nabla$ relative to this 
trivialization then (3.7), (3.8) and (3.9) show that the
transition function $S_j$ can be chosen so that it
has a holomorphic continuation into the exterior of $C_j$ 
in ${\bf P}^1$ (including $j=\infty$).  Each of the integrals
$$\int_{C_j}{{dx}\over {2\pi i}}\hbox{\rm Tr}\left(\partial_{\ell}
S_jS_j^{-1}\left(\partial_kS_jS_j^{-1}\right)'\right),$$
then vanishes by Cauchy's theorem.  This shows that if
$(t^0,\lambda^0)\notin\Theta$ then,
$$d\hbox{\rm Tr}\left(T_S^{-1}T_{dS}-T_{dSS^{-1}}\right)=0,\eqno 
(3.38)$$
at least for the special constructions associated with 
$(t^0,\lambda^0)=(t^1,\lambda^1).$ But we can now use (3.35) and (3.36)
to show that $\hbox{\rm Tr}\left(T_S^{-1}T_{dS}-T_{dSS^{-1}}\right
)$ is closed even 
without this restriction.   Define 
$$\omega_{\hbox{\rm JMU}}:=-\sum_j\hbox{\rm Res}_j\hbox{\rm Tr}\left
(\hat\alpha_j^{-1}\hat\alpha_j'dH_j\right),$$
which is the one form introduced by Jimbo, Miwa
and Ueno in [12].
Then (3.38), (3.35) and (3.36) together show that
$$d\omega_{\hbox{\rm JMU}}=0,\eqno (3.39)$$
for $(t,\lambda )\notin\Theta$.  This result and (3.36) (which is not 
restricted to the special choice $(t^0,\lambda^0)=(t^1,\lambda^1)$) then 
show that the right hand side of (3.35) is closed in 
general and we conclude that  $\hbox{\rm Tr}\left(T_S^{-1}T_{dS}-
T_{dSS^{-1}}\right)$ is
closed even when $(t^1,\lambda^1)$ is different from $(t^0,\lambda^
0).$
We are now prepared to state the principal result of 
this paper.
\vskip.1in\par\noindent
{\bf Theorem} 3.40.
{\sl\ Suppose that $\bar{\nabla}^0$ is a connection on the trivial 
bundle, ${\bf P}^1\times {\bf C}^p\rightarrow {\bf P}^1$, with simple type $
r_j$ singularities at
the distinct points
$a_j\in {\bf P}^1$ for $j\in \{1,2,\ldots ,n,\infty \}$ with $a_{
\infty}:=\infty$.  Let $(E,\nabla )$ 
denote the vector bundle with connection constructed as
a deformation of $\bar{\nabla}^0$ in Theorem 2.9.  Let $\Theta$ denote the
set of $(t,\lambda )\in {\cal D}$ such that $E|_{^{{\bf P}^1\times 
\{(t,\lambda )\}}}$ is not trivial.  For
$(t,\lambda )\notin\Theta$ let $\hat{\alpha}_j$ and $\alpha_j$ be defined as in (3.24) and (3.28).
Then
\vskip.1in\par\noindent
(i) The form defined by
$$\omega_{\hbox{\rm JMU}}=-\sum_j\hbox{\rm Res}_j\hbox{\rm Tr}\left
(\hat\alpha_j^{-1}\hat\alpha_j'dH_j\right),\eqno (3.41)$$
is a closed one form on ${\cal D}\backslash\Theta$ and there exists a 
holomorphic function $\tau_{\hbox{\rm JMU}}$ on ${\cal D}$ such that
$$\omega_{\hbox{\rm JMU}}=d\log\tau_{\hbox{\rm JMU}}.$$
(ii) The point $(t,\lambda )\in {\cal D}$ is a zero of $\tau_{\hbox{\rm JMU}}$ if and only
if $(t,\lambda )\in\Theta$. }
\vskip.1in\par\noindent
Proof.  Let $S$ denote the transition function defined
by (3.14) and (3.15).  Then as in the proof of Theorem
3.0 one can find an invertible holomorphic parametrix,
$$W\ni (t,\lambda )\rightarrow q(t,\lambda ),$$
so that 
$$T_{S(t,\lambda )}q(t,\lambda )^{-1}=I+\hbox{\rm trace class},$$
for $(t,\lambda )\in W.$  Define
$$\tau_q(t,\lambda ):=\det\left(T_{S(t,\lambda )}q(t,\lambda )^{-
1}\right).$$
Then it is clear that $\tau_q(t,\lambda )=0$ if and only if $(t,\lambda 
)\in\Theta .$
As above write $d=d_t+d_{\lambda}$.  Then the usual formula for
the derivative of a determinant gives
$$d\log\tau_q=\hbox{\rm Tr}\left(T_S^{-1}T_{dS}-dqq^{-1}\right),\eqno 
(3.42)$$
off the singular set $\Theta$.  Comparing this with $\omega$ defined 
by (3.18) above we find that:
$$\omega -d\log\tau_q=\hbox{\rm Tr}\left(dqq^{-1}-T_{dSS^{-1}}\right
).\eqno (3.43)$$
The left hand side is a closed form on $W\backslash\Theta$ and the 
right hand side is holomorphic on $W$.  Thus the right 
hand side of (3.43) is a closed form on $W$.  Since $W$ is
simply connected it follows that there exists a 
holomorphic function $\phi$ on $W$ so that
$$\omega -d\log\tau_q=d\phi .\eqno (3.44)$$
Consulting the definition (3.41) for $\omega_{\hbox{\rm JMU}}$ and the 
result (3.35) for $\omega$ one finds
$$\omega_{\hbox{\rm JMU}}-\omega =-\sum_j\hbox{\rm Res}_j\hbox{\rm Tr}\left
(\hat\alpha_j(loc)^{-1}\hat\alpha_j'(loc)^{-1}dH_j\right)\eqno (3
.45)$$
The right hand side of (3.45) is a holomorphic one form 
on $W$ and by (3.36) it is closed.
  Thus there exists a holomorphic function
$\varphi$ on $W$ so that 
$$\omega_{\hbox{\rm JMU}}-\omega =d\varphi .\eqno (3.46)$$
Adding (3.44) and (3.46) and writing $\Phi =\phi +\varphi$ we find
$$\omega_{\hbox{\rm JMU}}=d\log\tau_q+d\Phi =d\log\left(e^{\Phi}\tau_
q\right).\eqno (3.47)$$
Thus for some constant, $c,$ we have
$$\omega_{\hbox{\rm JMU}}=ce^{\Phi}\tau_q.$$
From this it follows immediately that $\omega_{\hbox{\rm JMU}}(t,
\lambda )=0$
if and only if $(t,\lambda )\in\Theta$ and this finishes the proof of the
theorem. QED

\centerline{\bf Acknowlegements}
The author would like to thank Doug Pickrell for helpful conversations.
This research was partially funded by NSF grant DMS 9401594.

\vskip.2in \par\noindent
\centerline{\bf References}
\vskip.1in\par\noindent
[1] Anosov, D.V., Bolibruch, A.A., {\sl The Riemann-Hilbert
problem, Aspects of Mathematics}: E; Vol. 22 (1994).
\vskip.1in\par\noindent
[2] Babbitt, D.G., and Varadarajan, V.S., {\sl Local moduli for
meromorphic
differential equations }, Ast\'erisque {\bf 169-170} (1989) p.1-217.
\vskip.1in\par\noindent
[3] Balser, W., Jurkat, W., and Lutz, D., {\sl A general theory of
invariants for meromorphic differential equations; Part II,  Proper 
invariants}, Funkciala Ekvacioj {\bf 22},(1979) p.257-283.
 \vskip.1in\par\noindent
[4] Bolibruch, A.A., {\sl The Riemann-Hilbert problem}, Russian
Math. Surveys, {\bf 45:2} (1990) p.1-47.
\vskip.1in\par\noindent
[5] Birkhoff, G.D., {\sl The generalized Riemann problem for linear
differential equations}, Proc. Amer. Acad. Arts Sc. {\bf 49 }
(1913) p.531-568.
\vskip.1in\par\noindent
[6] Deligne, P., {\sl Equations differ\'entielles \`a points singulieres
r\'eguliers}, Springer Lect. Notes, {\bf 163}, (Springer 1970).
\vskip.1in\par\noindent
[7] Faddell, E., Neuwirth, L., {\sl Configuration spaces}, Math.
Scand. {\bf 10}, (1962), p.111-118.
\vskip.2in \par\noindent
[8] Flaschka, H., Newell, A., {\sl Monodromy and spectrum
preserving deformations I}, Comm. Math. Phys. {\bf 76}, (1980)
p.65-116.
\vskip.1in\par\noindent
[9] Grauert, H. and Remmert, R., {\sl Theory of Stein Spaces},
(Springer, Berlin, 1979)
\vskip.1in\par\noindent
[10] Grauert, H., {\sl Analytische Faserungen \"uber
holomorph-vollst\"andigen R\"aumen}, Math. Annalen Band {\bf 135 }
(1958), p.263-273.
\vskip.1in\par\noindent
[11] Helmink, G., {\sl Deformations of connections, the
Riemann-Hilbert problem and $\tau -\hbox{\rm functions}$}, p.75-89 in:
Computational and Combinatorial Methods in System 
Theory, C.I. Byrnes and A. Lindquist (editors), Elsevier 
Science Publishers B.V. (North Holland, 1986).
\vskip.1in\par\noindent
[12] Jimbo, M., Miwa, T., and Ueno, K., {\sl Monodromy
preserving deformations of linear ordinary differential
equations with rational coefficients I}, General theory and
tau function, Phys. 2D, {\bf 2} (1981) p.306-352.
\vskip.1in\par\noindent
[13] Jimbo, M., Miwa, T., {\sl Monodromy preserving deformations
of linear ordinary differential equations II}, Physica D {\bf 2 }
(1981) p.407-448.
\vskip.1in\par\noindent
[14] Jimbo, M., Miwa, T., {\sl Monodromy preserving deformations
of linear ordinary differential equations III}, Physica D {\bf 4 }
(1983) p.26-46.
\vskip.1in\par\noindent
[15] Miwa, T., {\sl Painlev\'e property of monodromy preserving
equations and the analyticity of $\tau -\hbox{\rm function}$}, Publ.
R.I.M.S. Kyoto University {\bf 17-2} (1981) p.703-721.
\vskip.1in\par\noindent
[16] Malgrange, B., {\sl La classification des connections
irr\'eguli\'eres \`a une variable}, p.381-400, in:
Boutet de Monvel, L., Douady, A. and Verdier, J-L., 
Math\'ematique et Physique, Progress in Mathematics, {\bf 37},
(Birkh\"auser, Boston, 1983).
\vskip.1in\par\noindent
[17] Malgrange, B., {\sl Sur les d\'eformations isomonodromiques, I.
Singularit\'es r\'egu\-li\`eres}, p.401-426, in:
Boutet de Monvel, L., Douady, A. and Verdier, J-L., 
Math\'ema\-tique et Physique, Progress in Mathematics, {\bf 37},
(Birkh\"auser, Boston, 1983).
\vskip.1in\par\noindent
[18] Malgrange, B., {\sl Sur les d\'eformations isomonodromiques, II.
Singularit\'es irr\'egu\-li\`eres}, p.427-438, in
Boutet de Monvel, L., Douady, A. and Verdier, J-L., 
Math\'ema\-tique et Physique, Progress in Mathematics, {\bf 37},
(Birkh\"auser, Boston, 1983).
\vskip.1in\par\noindent
[19] Schlesinger, L., {\sl \"Uber eine klasse von differentialsystem
beliebiger ordnung mit festen kritischen punkten}, J.
Reine u. Angewendte Math., {\bf 141}, (1912) p.96-145.
\vskip.1in\par\noindent
[20] Sibuya, Y., {\sl Stokes' Phenomena}, Bull. Amer. Math. Soc.
83-5 (1977) p.1075-1077.
\vskip.1in\par\noindent
[21] Sibuya, Y., {\sl Perturbation of linear ordinary differential
equations at irregular singular points}, Funk. Ekvacioj {\bf 11}
(1968) p.235-246.
\vskip.1in\par\noindent
[22] Ueno, K., {\sl  Monodromy preserving deformations of linear
differential equations with irregular singular points},
R.I.M.S. Preprint 301 (1979)
\vskip.1in\par\noindent
[23] Varadarajan, V.S., {\sl Linear meromorphic differential equations:
a modern point of view}, Bull. Amer. Math. Soc. {\bf 33-1} (1996) 
p.1-41. \vskip.1in\par\noindent
[24] Wasow, W., {\sl Asymptotic expansions for ordinary
differential equations}, Interscience Publ. (1965)

\bye